\documentclass[fleqn,usenatbib]{mnras}

\usepackage{newtxtext,newtxmath}
\usepackage[T1]{fontenc}
\usepackage[utf8]{inputenc}
\usepackage{graphicx}
\usepackage{tikz}
\usepackage{xcolor}
\usepackage[percent]{overpic} 
\usepackage{subcaption}
\title[Radio emission in wide-orbit brown dwarfs]{Searching for radio emission from stellar wind-magnetosphere interaction or co-rotation breakdown in brown dwarfs}

\newcounter{equalcontrib}

\author[R. Pirvu-Malanda et al.]{
Rebeca Pirvu-Malanda$^{1}$%
\thanks{E-mail: rpirvu@ice.csic.es}%
\thanks{These authors contributed equally to this work.},
\setcounter{equalcontrib}{\value{footnote}} 
Simranpreet Kaur$^{1,2}$\footnotemark[\value{equalcontrib}], Daniele Vigan\`o$^{1,2,3}$\footnotemark[\value{equalcontrib}], 
\`Oscar Morata$^{1}$,
\newauthor
Josep Miquel Girart$^{1,2}$,
\'Alvaro S\'anchez-Monge$^{1,2}$, 
Devojyoti Kansabanik$^{4,5}$, 
Mario Damasso$^{6}$, 
\newauthor
Mayank Narang$^{7}$,
Juan Carlos Morales$^{1,2}$,
Gemma Busquet$^{2,8,9}$, 
Fabio Del Sordo$^{10,11}$,
Kaustubh Hakim$^{12,13}$, 
\newauthor
Manoj Purvankara$^{14}$
\\
$^{1}$Institute of Space Sciences (ICE-CSIC), Campus UAB, Carrer de Can Magrans s/n, 08193, Barcelona, Catalonia, Spain\\
$^{2}$Institut d'Estudis Espacials de Catalunya (IEEC), Esteve Terradas, 1, edifici RDIT,\\
Campus Baix Llobregat - UPC, 08860 Castelldefels (Barcelona), Catalonia, Spain\\
$^{3}$Institute of Applied Computing \& Community Code (IAC3), 07122, Palma de Mallorca, Spain\\
$^{4}$Cooperative Programs for the Advancement of Earth System Science (CPAESS), UCAR, Boulder, CO, USA\\
$^{5}$The Johns Hopkins University Applied Physics Laboratory, Laurel, MD, USA\\
$^{6}$INAF -- Osservatorio Astrofisico di Torino, Via Osservatorio 20, I-10025 Pino Torinese, Italy\\
$^{7}$Academia Sinica Institute of Astronomy and Astrophysics, 11F of Astronomy-Mathematics Building, \\
AS/NTU, No. 1, Sec. 4, Roosevelt Rd, Taipei 106216, Taiwan\\
$^{8}$Departament de Física Quàntica i Astrofísica, Universitat de Barcelona, Martí i Franquès 1, E-08028 Barcelona, Catalonia, Spain\\
$^{9}$Institut de Ciències del Cosmos (ICCUB), Universitat de Barcelona, Martí i Franquès 1, E-08028 Barcelona, Catalonia, Spain\\
$^{10}$ INAF, Osservatorio Astrofisico di Catania, via Santa Sofia, 78 Catania, I-95123, Italy\\
$^{11}$ Scuola Normale Superiore, Piazza dei Cavalieri, 7, 56126 Pisa, Italy\\
$^{12}$Institute of Astronomy, KU Leuven, Celestijnenlaan 200D, 3001 Leuven, Belgium\\
$^{13}$Royal Observatory of Belgium, Ringlaan 3, 1180 Brussels, Belgium\\
$^{14}$Department of Astronomy \& Astrophysics, Tata Institute of Fundamental Research, Homi Bhabha Road, Colaba, Mumbai-400005, India
}

\date{Accepted XXX. Received YYY; in original form ZZZ}
\pubyear{2025}

\begin{document}
\label{firstpage}
\pagerange{\pageref{firstpage}--\pageref{lastpage}}
\maketitle


\begin{abstract}
With the improvements in radio interferometry sensitivity, the quest for coherent radio emission from exoplanets and ultra-cool dwarfs, which is indicative of their magnetic fields, has gained significant momentum in recent years. We investigated the relatively unexplored possibility of radio emission from wide-orbit brown dwarf companions, which may radiate through rapid rotation, as in isolated ultra-cool dwarfs, or via interactions between their extended magnetospheres and the host star's wind. We analysed $\sim 60$ hours of Upgraded Giant Metrewave Radio Telescope and Karl G. Jansky Very Large Array data for a set of well-characterized systems previously unobserved at 0.3-2 GHz.
The targets include companions orbiting the G-type stars HD 26161 and BD-004475, the K-type HD 153557A and $\nu$ Oph, and the M dwarfs GJ 3626 and 2MJ01225093-2439505. 
No detections were obtained with 3$\sigma$ upper limits down to $\sim 25\,\mu$Jy/beam in Stokes V in the best cases. The light-curve analysis also revealed no evidence of short ($\gtrsim$ minutes), intense ($\gtrsim$ mJy) radio bursts. The upper limits provide tentative constraints on model parameters. However, the effects of model uncertainties, limited observational coverage, and intrinsic variability or beaming of the emission must be considered. The improvement in sensitivity of the next-generation radio interferometers will likely allow to go below the expected flux range over a much larger range of free parameters. 
\end{abstract}


\section{Introduction}\label{sec:intro}

The sample of radio-bright main-sequence stars has been quickly increasing in the last few decades thanks to more sensitive and extended surveys (e.g., \citealt{pritchard21,yiu24,driessen24}). This has also been accompanied by a better radio characterization of brown dwarfs, while exoplanets remain elusive. One of the main interests in their radio emission is its tight connection to their magnetism, which is still elusive: the use of Zeeman-Doppler imaging in these cold objects is impractical, due to the broadening of the available spectral lines (e.g., \citealt{reiners07}). In these cases, the potentially most direct probe of magnetic fields is the radio emission generated by the electron cyclotron maser (ECM, \citealt{melrose82,treumann06}), detected in the most magnetized planets in the solar system (e.g., \citealt{desch84,zarka92,zarka98,zarka07,zarka21}), brown dwarfs (e.g., \citealt{pineda17}), low-mass stars (e.g. \citealt{hallinan08,mclean11,villadsen19,vedantham20,vedantham21,callingham21,bastian22,zhang23,bloot24,kaur24,kaur25}), and hot magnetised stars (e.g. \citealt{trigilio00,leto20,leto20l}). ECM emission is coherent, circularly polarised, and highly beamed along a thin conic surface almost perpendicular to the local magnetic field. It is emitted mostly along X-mode waves at the fundamental electron gyro-frequency $\nu_c = eB/(2\pi m_e) = 2.8~B$ [G] MHz, where $B$ is the local magnetic field intensity, $e$ and $m_e$ are the electronic charge and mass. ECM arises in the presence of a distribution of non-thermal electrons (accelerated via e.g., reconnection, Alfv\'en waves..., \citealt{zarka98}), embedded in a strongly magnetised, tenuous plasma (such that the plasma frequency $\nu_p<\nu_c$).

\subsection{Planetary aurorae}

Within the solar system, the well-measured Jupiter's magnetic field \citep{connerney22} is the only one strong enough to induce ECM radio emission at frequencies $\nu_c$ above the Earth's ionospheric cut-off of $\sim$10 MHz. The energy budget for the Jovian radio emission is a combination of three energy sources: (i) planetary fast rotation, (ii) interaction with the solar wind, and (iii) interaction with the Galilean satellites, Io in particular (e.g., \citealt{zarka98,zarka21}). The brightness and rich phenomenology of the Jovian radio emission makes it the primary reference for the so far  radio-undetected extrasolar analogues.
The search for exoplanetary radio emission (see \citealt{callingham24} for a recent review) over the past two decades mostly relied on considerable expectations (e.g., \citealt{farrell99, griessmeier05, griessmeier07, zarka07}) for nearby hot Jupiters (HJs), which are gas giants orbiting very close to their host stars (see \citealt{fortney21} for a comprehensive review of their properties).
HJs are generally considered promising targets because their proximity to the star increases the expected emission via two different scenarios, corresponding to the above-mentioned Jovian energy sources (iii) and (ii), respectively: magnetic sub-Alfv\'enic star-planet interaction (SPI) causing aurorae close to the surface of strongly magnetised host stars (e.g. \citealt{lanza13,saur13,strugarek16,turnpenney18,kavanagh21}), and wind-magnetosphere interaction triggering aurorae over the planet itself, as it happens on Earth \citep{zarka01,zarka07,stevens05,griessmeier07}.  Another motivation was the expectation of strong magnetic fields on these objects, which however is an open question, as discussed below. Rotation is instead not expected to play a major role, because it is relatively slow, locked to the orbital period (days).

Early observational efforts targeting promising HJs only set loose upper limits of about 10–100 mJy/beam at different sub-GHz frequencies (74, 154, 333, and 1465 MHz, \citealt{bastian00, lazio04, murphy15}). With advanced upgrades in interferometric facilities, many targeted campaigns have progressively lowered the upper limits below mJy/beam \citep{lecavelier09,lecavelier11,lynch17,lynch18,degasperin20,turner21,turner24,narang21,narang21b,narang23,narang23b,narang24,narang24b,ortiz24,route23,route24,shiohira24}. Satellite interaction has been the motivation for recent efforts to target exo-moon candidates, mostly around hot Saturns, again with no detection \citep{narang23, narang23b}.

Overall, most promising radio targets for sub-Alfv\'enic SPI from short-period planets (not necessarily giants) around highly magnetised stars have resulted in non-detections (see e.g. \citealt{narang24,pena25}), but interesting signals have been revealed from the planet-hosting M dwarfs Proxima Centauri \citep{zic20, perez21} and YZ Ceti \citep{pineda23}, for which follow-up campaigns will elucidate possible modulation with the  orbital period of their planets. Note that ECM from sub-Alfv\'enic SPI originates close to the star, so  that the frequency depends on the stellar magnetic field,  while the dependence of the expected radio flux with the companion's field is highly model dependent \citep{cauley19}. Although potentially relevant in many systems and interesting for many reasons, sub-Alfv\'enic SPI does not apply to the wide-separation targets presented in this work and will not be discussed further.

On the other hand, the only reported claims for direct exoplanetary radio detections (i.e., originating close to the planetary surface like in solar planets) include one LOFAR detection of short radio bursts from the $\tau$ Bo\"otis system \citep{turner21}, which has not been confirmed in later follow-up observations \citep{turner24,cordun25}, a possible hint of emission from HAT-P-11  system seen with GMRT \citep{lecavelier13}, which could be interpreted as ECM with a local magnetic field of about 50 G, and the recent highly circularly polarized burst from HD 187399 system, detected by NenuFAR \citep{zhang25}.
In all cases, a crucial challenge is that confirming the exoplanetary origin of a radio detection, rather than the stellar one, essentially implies proving a modulation with the planetary orbit  or rotation (which coincide for tidally locked HJs), which requires a good phase coverage. 

\subsection{Radio-loud ultracool dwarfs}

Much more successful has been the characterization of radio emission from isolated ultracool dwarfs (UCD, spectral class $\geq$ M7). The non-thermal radio emission for UCDs can have two components: weakly polarized quiescent emission often attributed to incoherent gyro-synchrotron from Jovian-like radiation belts (e.g. \citealt{berger01,berger02,berger06,loinard07,burgasser15,climent23,kao23}), and beamed, more variable, highly circularly polarized coherent emission, attributed to ECM (e.g., \citealt{route12,route13,williams15,williams17,route16,lynch16,pineda17,kao16,kao18,vedantham20bd,rose23}). Many UCDs present both components \citep{route12,pineda17}.  The ECM variability is expected from the combination of beaming, geometry, and intrinsic variability of magnetospheric plasma dynamics. ECM bursts are often modulated with the rotation and can constrain the magnetic configuration \citep{lynch15}. The combination of geometrical and magnetic configurations can also allow detection of auroral emission from both poles, so that the received flux has lower net polarization and looks more quiescent \citep{williams17}. Very Large Baseline Interferometry (VLBI) observations are probably the most direct way to disentangle the radiation belt and auroral contributions to the quiescent emission \citep{kao23,climent23}, but this might be possible only for the closest  radio-loud UCDs.

Multi-wavelength studies of activity indicators and radio emission from cool dwarfs show that there is a clear shift, across the L spectral type, from a Sun-like chromospheric/coronal paradigm, to a planetary-like auroral scenario, based on the circulation of field-aligned large-scale magnetospheric currents \citep{pineda17}. Therefore, in isolated brown dwarfs, coherent radio emission is typically associated with aurorae \citep{hallinan07,rodriguezbarrera15}, powered by their fast rotation \citep{mclean12}, like in Jupiter. In fact, UCDs are seen to spin with a frequency going from solar giant planets' values ($\sim$10 h), down to extremely fast rotations of $\sim$1 h \citep{tannock21}. Under these conditions, the key engine behind aurorae is thought to be the co-rotation breakdown of magnetospheric plasma \citep{cowley01}, a mechanism initially proposed for planets. Notably, rotation acts as the driving engine behind the incoherent gyrosynchrotron and ECM emissions in hot magnetic stars as well. However, unlike UCDs and planets, these stars have an abundant plasma supply originating from their own stellar winds. Apart from the difference in the plasma supply, the basic electrodynamic engine governing the centrifugal break-out and associated radio emission is indeed possibly the same across a wide range of objects, from hot stars to planets \citep{leto21,owocki22}, as long as their rotation is fast and they have a large (i.e., not compressed) magnetosphere.

Volume-limited surveys and targeted campaigns generally agree on a $\lesssim 10\%$ occurrence rate of radio detection (at several GHz) for UCDs \citep{route13,williams17,rose23,kao24,callingham24}, with  fast-rotating (period $P_{\rm rot}\lesssim 5$ h, with rare exceptions, \citealt{magaudda24}) and H$\alpha$-emitting UCDs \citep{kao16} being much more likely to emit in radio. 
The derived occurrence rate could derive from a combination of low duty cycles due to beaming (with some exceptions, \citealt{rose23}), limited phase coverage, and, more speculatively, the  presence/absence of plasma-providing satellites, which is difficult to test given the major challenge of confirming exo-moon signals even around main-sequence stars (e.g., \citealt{kipping25}). 
Note also that, within the UCD spectral classes, few T dwarfs (\citealt{kao16,kao18,pineda17,vedantham20,rose23,vedantham23}, the latter being a T-dwarf binary) and no Y dwarfs \citep{kao19} have been detected. Part of the reason could be the lower magnetic field expected for colder/less massive objects, according to the widely used scaling laws  (\citealt{christensen09}, see below), which, on the other hand, underestimate the surprisingly high field (few kG)  inferred from the ECM emission in a few UCDs \citep{kao16,kao18}. Importantly, estimating the radio-loud UCD occurrence rate at lower frequencies ($\lesssim$ GHz), more sensitive to lower magnetic fields, is more challenging due to the scarcity of observations (see e.g. \citealt{zic19}).

\subsection{Aims of this study}

We present an observational campaign with the upgraded Giant Metrewave Telescope (uGMRT) and Karl G. Jansky Very Large Array (JVLA), consisting of $\sim$ 60 hours of dedicated deep-field observations of six systems likely containing brown dwarfs, orbiting at large separations from their host stars. These systems were previously not targeted with dedicated observations, at least in the $\sim 0.3-2$ GHz range presented here.

The choice of the targets is complementary to most of the studies mentioned above (isolated UCDs or closely orbiting planets). However, it was already briefly mentioned in \cite{farrell99}, who estimated the expected flux in a few brown dwarfs closely orbiting main-sequence stars. Radio emission from wide-orbit brown dwarfs can potentially be powered by any of the three channels mentioned above: fast rotation, wind-magnetosphere interaction, and the presence of satellites.

Regarding the first one, wide-orbit brown dwarfs are expected to spin at periods similar to the ones of isolated UCDs (few hours, \citealt{pineda17,tannock21}) and solar giants, due to the absence of relevant spin-down torques like tidal forces or winds (see \citealt{batygin18} for the similar rotational evolution of cold Jupiters), thus making co-rotation breakdown and its associated radio emission possible. Moreover, given the expected fast rotation, our repeated few hours-long observations likely cover a large part of the rotational phases, which is essential in the presence of expected rotational modulation.

Given the available observables, the expected radio flux can be quantitatively estimated for the wind-magnetospheric interaction \citep{stevens05,griessmeier07} and for models accounting for the rotation \citep{farrell99} and Jovian-like magnetospheric currents \citep{hill01}. The former relies essentially on the radiometric Bode's law and radio-magnetic law,  phenomenological correlations seen among the five most magnetised solar planets between the radio power and the incident solar wind kinetic and magnetic flux, respectively (e.g., \citealt{desch84,zarka98,farrell99,zarka01,lazio04,callingham24}). Since the emission frequency scales with the magnetic field intensity, which in turn is thought to depend mostly on the mass (see below), such models indicate massive objects as promising targets for radio emission. For this reason, mass is our main selection criterion. Moreover,\citep{nichols16} showed that, when accounting for magnetospheric electrodynamics in more detail, those phenomenological models have been shown to overestimate the radio power for HJs (which would need confirmed radio detection for a clear answer) and underestimate it for more distant companions, which is an additional reason to focus on distant companions. Finally, binarity has been proposed to enhance the coherent radio emission in UCDs \citep{kao22,vedantham23}, possibly due to the mutual feeding of magnetospheric plasma and energy via winds, perhaps analogous to the wind-magnetospheric one here considered.

 In principle, theoretical radio flux estimates could be used as educated guesses to select the best targets. In practice, the lack of data beyond the solar planets, the huge uncertainty on the underlying parameters and the intrinsic expected variability, together with the mismatch of non-detection of tens of campaigns targetting exoplanets, put a word of caution on the reliability of the models, which will only be considered for an a-posteriori interpretation of the results.
 

Besides these physical motivations, note that very few studies have aimed at finding radio emission from wide-orbit companions, resulting in upper limits: \cite{cendes22} looked at several directly imaged massive exoplanets at 4-8 GHz, while \cite{shiohira24} targeted the $\beta$ Pictoris system at 250-500 MHz. Finally, most detections of isolated UCDs have been conducted at a few GHz, higher than the frequency range here explored, with a few exceptions of MeerKAT \citep{rose23}, GMRT \citep{zic19}, and LOFAR \citep{vedantham20bd,vedantham23} observations. 

In Sect.~\ref{sec:selection}, we discuss the targets, including selection criteria, properties, and flux estimates. In Sect.~\ref{sec:obs}, we present the observed sample, and the data analysis. The obtained upper limits and the comparison with different theoretical models are shown in Sect. \ref{sec:results} and discussed in Sect.~\ref{sec:Discussion}.

\section{Target selection}\label{sec:selection}

For the target selection, we have looked for sub-stellar objects with: (i) masses $M>10~M_{\rm J}$ (where $M_{\rm J}$ is the Jupiter mass) since they are expected to be highly magnetised (see next sub-section), (ii) distance within 50 pc, (iii) observable by uGMRT and VLA (dec $>-45^\circ$), and (iv) no previous dedicated observations by VLA or uGMRT in the public archives, at the date of the proposals. Note that in some cases only the minimum mass is available, $M\sin i$, where $i$ is the inclination angle. In that case, the estimated magnetic field, and the associated cyclotron frequency, is then a lower limit. When filtering the NASA exoplanet archive with criteria (i), (ii) and (iii), we obtain about 40 systems (we would obtain roughly the double of them if considering $M>5~M_{\rm J}$), about half of which have been targeted by previous radio campaign with uGMRT and/or VLA. Most of the already observed objects, excluded by criterion (iv), have more narrow orbits: they are naturally very interesting, since they can interact more with the host star and provide more radio flux, but existing dedicated observations and studies are present, so that we neglect them.
Note that we also included other systems, currently removed or not appearing in the exoplanetary archive (as for December 2025) due to the large masses. The reasons behind these criteria and the final targets are detailed in the rest of the section.

\subsection{Frequency range and magnetic field}\label{sec:freq_mf}

We aim at observing targets that can potentially have magnetic field strengths of a few hundred gauss, corresponding to a fundamental ECM frequency $\nu_c\sim 0.3-2$ GHz, where we carry our observations.
In order to estimate the magnetic field strength of the brown dwarf at the polar surface, $B$, we rely on the simplest version of the available scaling laws \citep{Arge1995, stevens05, farrell99}, which essentially predict a proportionality between the dipolar component of the surface magnetic field $B$ (at the pole) and the mass-radius ratio:
\begin{equation}\label{eq:magnetic_scaling}
    B = B_{\rm J}\frac{M}{M_{\rm J}}\frac{R_{\rm J}}{R},
\end{equation}
scaled to the surface polar Jovian values (we take $B_{\rm J}\sim 8.6$ G, see the dipolar component from \citealt{connerney18,connerney22}). Since the radius $R$ is quite a flat function of $M$ for the gas giants-brown dwarf mass range (e.g., \citealt{chabrier00,canas22} and App. A of \citealt{griessmeier07}), in this work we assume $R=1\,R_{\rm J}$ for simplicity.
As discussed in e.g. \cite{farrell99, stevens05, griessmeier07}, different, more elaborated scaling laws exist, which differ from each other in slightly different power-law dependencies with mass and radius, but roughly keep the same expected overall trend for solar planets. The most well-known scaling law is from \cite{christensen09}, which combined dynamo simulations of fully convective objects and the measured magnetic fields of the Earth, Jupiter and rapidly rotating low-mass stars with observed magnetic fields. That scaling law, valid for fast rotators (see e.g. \citealt{reiners08} for a discussion about rotation and activity in stars), predicts that the factor deriving the magnetic field is the convective heat flux in the dynamo region. \cite{reiners09} and \cite{reiners10} reformulate it in terms of more directly observable quantities, to obtain a relation, calibrated with stars, of the form $B\propto L^{1/3} M^{1/6} R^{-7/6}$, where $L$ is the internal luminosity, which is unknown in most cases here considered. If we assume it to scale as $L\sim M^\alpha$, with $\alpha\lesssim 3$ like in low-mass stars, the overall dependence of $B$ with both $M$ and $R$ is only slightly steeper than the simple eq.~(\ref{eq:magnetic_scaling}) used in this work. Moreover, the \cite{reiners09} scaling, originally calibrated for low-mass main-sequence stars, may not be directly applicable to sub-stellar objects (see also Appendix A of \citealt{vigano25} for a detailed discussion). In the case of brown dwarfs, this scaling law appears to underestimate the magnetic field strengths derived from ECM observations of a few UCDs \citep{kao16,kao18}. For exoplanets, the situation is even less certain, since the only available estimates are indirect. These have been proposed for a few hot Jupiters, where possible magnetic modulation of stellar activity with orbital phase has been observed \citep{cauley19}. The inferred field strengths, around $B \sim \mathcal{O}(10^2)$ G, are roughly consistent with some theoretical predictions based on the same scaling law \citep{yadav17,kilmetis24}, but they are highly model-dependent. However, other HJ evolutionary models, using the same scaling laws, consider the suppression of convection and dynamo due to strong heat deposition in the outer layers, finding much lower, Jupiter-like magnetic fields \citep{elias25,vigano25}. \cite{zaghoo18} consider the conductivity and internal structure of gas giants and also predict $\sim$ gauss fields for both cold and hot Jupiters. Therefore, the expected range of HJ magnetic fields is an open question.

 
Qualitatively different scaling laws could apply to slow rotators. In that case, the dynamo is regulated by the spin frequency $\omega$, so that scaling laws of the form $B\sim M\omega/R$ have been proposed (see \citealt{sanchezlavega04,stevens05} and references within). However, the recent study by \cite{elias25} suggests that even HJs, which rotate much slower than their cold analogues due to tidal locking, are in the fast rotator regime. For our purposes, the time that the brown dwarf would need to synchronize its initial fast rotation to the orbital period ($P_{\rm orb}\gg 2\pi/\omega$) is given by \cite{griessmeier07,elias25} and references within, as:

\begin{equation}
\tau_{\rm syn}\sim  \frac{G \alpha}{36 \pi^4} \frac{M Q_P' P_{\rm orb}^4 \omega}{R_{\rm P}^3} \approx \left(\frac{M_{30} P_{yr}^4 \omega_{1}}{R_1^3}\right)1.2\cdot10^{7} {\rm \,Gyr} ~,
\label{eq:tau_sync}
\end{equation}

\noindent where we fixed the values of moment of inertia factor $\alpha=I/(MR^2)=0.4$ for simplicity (homogeneous sphere), the normalised tidal dissipation factor to $Q'_P = 5 \cdot10^5$ as in \cite{griessmeier07}, and we have defined $M_{30}=M/30\,M_{\rm J}$, $P_1=P_{\rm orb}/1{\rm \,yr}$, $R_1=R/R_{\rm J}$, and $\omega_1 = \omega/(2\pi/1 {\rm \,hr})$, i.e. normalizing $\omega$ to a period of 1 hour, the shortest found in UCDs, with much shorter periods being too close to breakup values \citep{batygin18}. This clearly shows that, for typical brown dwarf masses and orbital periods here considered (years to decades), the synchronization timescales are larger much than Hubble time: tidal forces can be safely neglected and the objects are likely to retain the post-formation fast rotation.
Therefore, one can expect wide-orbit brown dwarfs to clearly be fast rotators, making $B$ independent on the (unknown) rotation rate. 

In summary, it is important to keep in mind the high intrinsic uncertainties and dispersion ($\sim 60\%$ in \citealt{reiners09} scaling law). Taking into account all these considerations, we will use the simple eq.~(\ref{eq:magnetic_scaling}), which implies that the associated fundamental cyclotron frequency at the polar surface, $\nu_c^p \sim 470 \frac{M}{20~ M_{\rm J}}$~MHz, and its second harmonic $2\nu_c^p$, can be in the range of our interest ($\gtrsim 0.3$ GHz) for brown dwarfs.

\begin{table*}
\centering
\vspace{1pt} 
\begin{tabular}[h]{cccccccc}
\hline
Target & Host star & Age & Distance & Mass & Semi-major & RA & Dec \\
name$\nmid$ & spectral type & [Gyr] & [pc]  & [$M_{\rm J}$] &  axis [au] & [hh:mm:ss] & [dd:mm:ss] \\
\hline
      HD 153557 d & K3 V & - & $17.941 \pm0.008$ & $27.5\pm3.5$& $21.4^{+1.5}_{-1.7}$ & 16:57:52.96 & +47:22:04.28 \\
      2M0122-2439 B & M3.5 V & $0.12\pm0.01$ & $33.83_{-0.09}^{+0.09}$ & $24.5 \pm 2.5$& $52\pm6$ & 01:22:51.07 & -24:39:52.61 \\
      HD 26161 b$^\dagger$ & G0 V & - & $40.91\pm 0.08$ &  $58.65\pm 17.80$ & $23.2_{-8.0}^{+21.0}$ & 04:09:38.93 & +31:39:08.76 \\
      BD-00 4475 B &  G0 V & - & $42.64 \pm 0.18$  & $50.9^{+0.5}_{-0.7}$ & $1.64 \pm 0.05$  & 23:10:50.60 & +00:24:30.82 \\
      $\nu$~Oph b$^+$ & K0 III & $0.65\pm0.17$ & $46.21 \pm 1.73$  & $22.2$ & $1.79$ &17:59:01.58 & -09:46:36.88 \\
      $\nu$~Oph c$^+$ & " & " & " & $24.7$ & $5.93$&  " & " \\
      GJ 3626 B$^\dagger$ & M4.0 V & - & 22.8 & $34.5\pm 1$ &$\sim3.08$ &10:50:26.01 & +33:06:04.06 \\
      \hline
\end{tabular}
\caption{Properties of the selected targets. Columns indicate: name, spectral type of the host star, distance to Earth, quoted constraints on mass, semi-major axis of the orbit, estimated cyclotron frequency, and RA and Dec J2000 coordinates.\\
$\dagger$ For these objects, the mass column indicates in reality the minimum mass, $M\sin i$, due to the unconstrained inclination angle $i$.\\
$^+$ We take the values of mass and semi-major axis from \citep{quirrenbach19}, who did not report the associated errors, but that could be of the order of a few \% maximum, according the errors in their best solution, and from other references (see text).\\
$\nmid$ The lower/uppercase in companion's names follow the notation suggested by SIMBAD and recent literature, as for December 2025.}
\label{tab:Properties_of_the_targets}
\end{table*}

\subsection{Properties of the selected targets}\label{sec:sample}

Based on the aforementioned criteria, we have been proposing observations of several of the targets in the last few years. We present a list of the observed systems in Table~\ref{tab:Properties_of_the_targets}. The sample is quite heterogeneous in terms of stellar properties. However, all the candidates are massive with (minimum) masses ranging between $\sim 20~M_{\rm J}$ and $\sim 50~M_{\rm J}$. Separations to their host star range between $\sim$1.8 and $\sim$50 au. Below, we briefly describe the target systems, indicating in parentheses the short name used hereafter.

\subsubsection{HD 153557}

HD 153557 (HIP 83020) is the primary component of a binary system, which has also a wide K3V companion part. This wide triple system, ADS 10288 \citep{kiselev09,hirsch21}, is located at a distance of 17.9 pc. The binary is composed by the BY Draconis variable K3V \citep{loth98} and an M1.5V dwarf, HD 153557 B \citep{lepine13}. Not surprisingly for a chromospherically active star, it is X-ray bright (listed in NEXXUS-2 and 2RXS). \cite{Feng_2022} combined astrometry and radial velocity to detect, around HD 153557, two close Neptune-like planets (b, c), with orbital periods of $\sim$ 1 and 2 weeks, and a brown dwarf (d) with a mass $M\sim27.5\pm 3.5~M_{\rm J}$, and a slightly eccentric ($e\sim 0.15$) orbit with semi-major axis $a\sim 21.4$ au, which we take as nominal separation. The latter, being the most massive, is primary candidate of interest for potential radio emission in this study. \cite{nordstrom04} found $v_{\rm rot}\sin i\simeq6$ km/s, indicating relatively slow rotation, so that we don't expect $B_\star \gg B_\odot$. 

\subsubsection{2MASS J01225093-2439505 (2M0122-2439)}

This system lies 33 pc away and is composed at least by an M3.5V star and a L-type companion, dubbed also [PS78] 191B, that orbits at about 50 au, and that was direct imaged for the first time in 2013 by Keck II/NIRC at a 1.45” separation \citep{bowler13,hinkley15,sebastian21}, corresponding to a physical separation $a\sim 52\pm6$ au. \cite{bryan20} set constraints on the orbit, finding an a-posteriori distribution for the semi-major axis peaking at $\sim 30$ au, consistent with the projected separation, and very high obliquity.
There are several hints about the young age of the system, constrained to be between 20 Myr (lower limit provided by the lack of lithium signatures) and 120 Myr (assuming its membership to the AB Dor moving group, see \cite{bowler13} and references within). The system is X-ray bright, detected in both 2RXS and eROSITA, at a similar level, translating in a high mass-loss rate with our assumptions. The stellar period is also short, $1.49$ d (\citealt{doyle19}, compatible with the rotational velocity from \citealt{bowler13}), as expected given the age.
We assume the mass value of $M\sim 24.5~M_J$ and separation of $a \sim 52$ au from \cite{bowler13} (given these constraints and the very long orbit, the orbital displacement between its discovery and the radio observations is negligible). Importantly, a tentative evidence for a rotational photometric modulation of the companion, interpreted as rotation, has been observed, at $P_{\rm rot}=6$ h \citep{zhou19}, close to the average period of known isolated UCDs \citep{tannock21}, and possibly short enough to allow for co-rotation breakdown. The young age, which implies higher values of the educated guesses of $n_{\rm sw}$, $v_{\rm sw}$ and $B_\star$, and the fast rotation of the companion, makes this object particularly interesting. 

\subsubsection{HD 26161}
HD 26161 (HIP 19428) is a G0 V type star at $d \simeq 40$ pc, around which a sub-stellar companion was reported by \cite{rosenthal21} via radial velocity measurements, which are consistent with a highly eccentric orbit ($e\sim 0.82^{+0.06}_{-0.05}$) and a minimum mass of $M\sin i = 13.5^{+8.5}_{-3.7}~M_{\rm J}$. \cite{Feng_2022} combined radial velocity and astrometric data to report a statistically compatible value of eccentricity, but a higher mass though with large uncertainties, $M=28.5^{+20.0}_{-0.2}~M_J$ (but see \cite{lagrange23} for considerations about the limited phase coverage). \cite{an25} further refined the analysis, favouring $M =58.65 \pm 17.80~M_{\rm J}$, with a semi-major axis $a=23.2^{+21.0}_{-8.0}$ au. We use the latter as a separation because, within the limited phase coverage and considering the uncertainty on the orbit, we cannot quantify the separation at the time of the observations.

\subsubsection{BD-00 4475}
BD-004475 (HIP 114458) is a Sun-like, G0 V star, located at 42 pc from us. 
It is orbited by a brown dwarf, with a minimum mass initially reported to be $M\sin i = 25.05\pm2.23$
\citep{dalal21}. More recently, \cite{unger23} refined the solution and found a true mass of $M=50.9^{+0.5}_{-0.7}$, with a well-constrained eccentricity $e\sim 0.38$. Using this solution, we derive separations $a$ of $2.1711_{-0.0046}^{+0.0052}$ au for the observation on 21/11/2022, $1.561_{-0.017}^{+0.021}$ au on 5/3/2023, $1.431_{-0.019}^{+0.015}$ au on 26/5/2023, and $1.417_{-0.016}^{+0.013}$ au on 28/5/2023. Considering these numbers, we use an average separation of $a\sim 1.64$ au (which is, coincidentally, the inferred semi-major axis) to estimate the radio flux from wind-magnetosphere interaction. The reported slow stellar rotation ($v_{\rm rot}\sin i \sim 3$ km/s, \citealt{dalal21}) likely excludes values $B_\star \gg B_\odot$.

\subsubsection{$\nu$ Oph}
$\nu$ Oph (HIP 88084) is a giant K0 III star, with a mass about $2.8~ M_\odot$ \citep{jofre15}, with an age of $t\sim 0.65\pm 0.17$ Gyr. 
This star hosts two brown dwarfs that have been discovered by radial velocity measurements, with a $\sim 90^\circ$ inclination and small eccentricity: $\nu$ Oph b and $\nu$ Oph c  with masses of $22.2~M_{\rm J}$ and $24.7~M_{\rm J}$, respectively (\citealt{quirrenbach19}, see also \citealt{quirrenbach11,sato12,teng23}). Within our sample, this is the only evolved (i.e., post-main-sequence) star, for which the validity of the density and velocity prescriptions above is likely not applicable, we take the solar values as the fiducial ones. \cite{plachinda21} do not detect any statistically relevant magnetic field at the stellar surface, and the stellar rotation is probably very slow \citep{massarotti08}, thus excluding values of $B_\star \gtrsim B_\odot$.

\subsubsection{GJ 3626}
GJ 3626 (G 119-037) is a system, located at 22 pc, consisting of a M4.0V star and a companion, GJ 3626 B, discovered through CARMENES radial velocity measurements \citep{baroch21}, having a minimum mass $M\sin i \sim 34.5\pm 1 ~M_{\rm J}$, an $e\sim 0.43$ orbit, and a semi-major axis $\sim 3$ au. Using the orbital and inclination constraints by \cite{baroch21}, we estimate the separation at the time of observation to be roughly $a\sim 4$ au. The star is quite inactive, the rotational period is likely large \citep{baroch21}, thus excluding values $B_\star \gg B_\odot$. Its X-ray luminosity is accordingly low, as detected by XMM-Newton (appearing both in NEXXUS-2 and 4XMM), implying a mass-loss rate comparable to the Sun. Note that \cite{baroch21} favoured the brown dwarf scenario, but did not discard the possibility of being a secondary low-mass main-sequence star. If the latter was the case, it is anyway interesting to observe it in radio, given the relatively high number of radio-loud binary M dwarfs (e.g. \citealt{yiu24,kaur25}).



\begin{table*}
\centering
{\small
\centering
\vspace{0.5pt}
\begin{tabular}[h]{ccccccccccc}
\hline
Target & Date & Flux & Phase &Beam size & On-source & Band & $\sigma_I$ & $\sigma_V$ & $L_{\nu}^V$\\
(Proposal ID) & & Cal. & Cal.& [arcsec] & time [h] & [GHz] & [$\frac{\mu{\rm Jy}}{{\rm beam}}$] & [$\frac{\mu{\rm Jy}}{{\rm beam}}$] & [$\frac{{\rm erg}}{{\rm s~Hz}}$] \\
\vspace{.002cm}\\
\hline
\vspace{.002cm}\\
HD 153557 d & 03/11/2023 & 3C286 &1609+266 & $10.2\times3.6$ & 1.0 & 0.55-0.75  & 350 & 28\\  (uGMRT 45$\_$042)
& 01/12/2023 & & & $7.6\times4.1$ & 1.0&  & 320 & 35 \\ 
& 04/01/2024 & &  & $4.9\times3.4$ & 1.0 &   & 310 & 27 \\
& 15/02/2024  &  &  & $4.5\times3.9$ & 1.0 &  & 260 & 36 \\
& {\em combined}&&&&&&320 &24 & $2.76\cdot10^{13}$ \\
2M0122-2439 B & 13/11/2021 & 3C48 & 0025-260& $8.8\times7.8$ & 2.5 & 0.3-0.5 & 1578& 362 \\  (uGMRT 41$\_$007)
& 18/11/2021 & & & $8.9\times6.5$ & 2.5 && 275& 31\\
& 30/11/2021 & & & $8.0\times6.2$ & 2.5 && 480& 39\\
& 01/01/2022 & & & $8.7\times5.6$ & 2.5 & &570 & 53\\
& {\em combined}$^\dagger$ &&&&&& 370 & 30 & $1.23\cdot10^{14}$\\
HD 26161 b & 29/04/2023 & 3C138 & J0414+3418 & $4.9\times4.4$ & 0.85 &  1-2 & 17 & 14 \\  (VLA 23A-221)
& 27/05/2023 & & & $4.5\times4.1$ & 0.85 & & 23 & 21 \\
& {\em combined}&&&&&& 16 &8 & $4.81 \cdot10^{13}$ \\
BD-00 4475 B & 26/05/2023 & 3C48 & J2323-0317 & $5.7\times4.2$ & 0.85 & 1-2& 16 & 13 \\  (VLA 23A-221)
& 28/05/2023  & & & $5.8\times4.3$ & 0.85 & & 16 & 13\\
& {\em combined}&&&&&&16 &8 & $5.87\cdot10^{13}$\\
BD-00 4475 B  & 21/11/2022 & 3C48 & 2340+135 & $4.7\times3.6$ & 2.5 & 0.55-0.75 & 50 & 11\\ (uGMRT 43$\_$060)
& 05/03/2023 & & & $5.9\times3.5$ & 3.0 & & 45 & 13 \\
& {\em combined}&&&&&& 37 & 9 & $5.87\cdot10^{13}$\\
$\nu$ Oph b/c & 11/01/2023 & 3C286 & J1822-0938 & $9.9\times4.4$ & 0.85 & 1-2 & 20 & 16 \\  (VLA 23A-221)
& 11/05/2023  & & & $6.7\times4.3$ & 0.85 & & 20 & 15\\
& {\em combined}&&&&&&14 &9 &$6.90 \cdot10^{13}$ \\
$\nu$ Oph b/c & 26/11/2022 & 3C286/3C48 & 1822-096 & $5.1\times3.5$ & 4.9& 0.55-0.75& 33& 9\\ (uGMRT 43$\_$060)
& 14/03/2023 & & & $4.9\times4.3$ & 5.5 & & 62 & 20 \\
 & {\em combined}&&&&&&35&10 &$7.66\cdot10^{13}$ \\
GJ 3626 B & 11/06/2023 & 3C286 & 1021+219 & $3.8\times2.9$ &1.0&  0.55-0.95 & 72 & 68 \\ (uGMRT 44$\_$045) 
& 29/07/2023 & 3C147 &  3C147 & $5.1\times2.8$ & 1.0 & & 60 & 50 \\
& {\em combined} &&&&&&37&37 & $6.90\cdot10^{13}$\\
\hline
\end{tabular}
}
\caption{Characteristics of the observations. Columns indicate, respectively: short name of the target, date of the observation, proposal ID (the PI of each proposals was one among S. Kaur, R. Pirvu-Malanda and D. Vigan\`o in all cases), flux and phase calibrators, beam size, on-source time, usable band, background level of Stokes I and V, and 3$\sigma$ upper limit to the Stokes V isotropic luminosity (for the combined images only). When more than one observation was performed in the same band, we also reported the background values for the combined images.\\
$\dagger$ The combined image here includes only three datasets (18/11/2021, 30/11/2021, 01/01/2022), excluding the 13/11/2021 noisy observation.\\
$\ddagger$ This uGMRT Band 5 doesn't currently offer reliable calibration of circular polarization, since data are taken in the XY basis.}
\label{tab:obs}
\end{table*}

\section{Observations and analysis}\label{sec:obs}

Table \ref{tab:obs} shows the observation details for the targets. It includes a series of observations performed between 2021 November and February 2024 with either uGMRT or VLA, for a total observing time of about 60 h ($\sim$ 40 h on-source). Most of the estimated values of $\nu_c$ in the selected targets lie within the frequencies of the Band 4 of uGMRT (with usable band 0.55-0.75 GHz), which is what we mostly used. Some systems were observed with VLA in Band L (1-2 GHz): BD-00 4475 B because of its large mass, HD 26161 b because of the availability of a lower limit only for the mass and $\nu_c$ (see above). Moreover, for two sources, $\nu$ Oph b/c and BD-00 4475 B, having observations with both VLA and uGMRT allows a broader coverage in frequency, which allows also to explore the possibility of higher ECM harmonics or underestimated magnetic fields.

Given the expected irregular nature of radio emission, we observed each target between two and four times, with an on-source time from 1 to 8 hours, depending on the required time to theoretically achieve a noise level $\sigma_{\rm rms}\sim 10-20~\mu$Jy/beam in the clean images of each observation. In some cases, the resulting background noise was higher than the estimated one, usually due to the presence of bright sources in the field of view (see App.~\ref{app:images}), and, at a lesser extent, due to partial data loss resulting from RFI flagging.

The spectral setup of uGMRT consists of one spectral window with 2048 channels. The channels have a spectral resolution of 98 KHz. For VLA observations, the spectral setup consists of 14 spectral windows with 64 channels, and each channel has a spectral resolution of 1 MHz.

\subsection{Calibration and imaging}

The entire dataset was analysed using the Common Astronomy Software Applications (CASA) \citep{CASA}. VLA and uGMRT data were flagged and calibrated using the standard VLA CASA Calibration Pipeline and the CASA Pipeline-cum-Toolkit for uGMRT data Reduction (CAPTURE) pipeline \citep{kale21}, respectively. After an initial calibration was performed, the diagnostic plots were checked again to correct for remaining artifacts in the data.

Note that in several uGMRT observations, we found a butterfly-like trend for the frequency versus phase plots for the longest baselines in the phase calibrator scans. This prompted us to apply additional corrections for time delays for uGMRT data, similarly to what was done by \cite{kaur24}. Although this does not affect much the results, since we only have upper limits, we have included this additional correction step in the pipeline. In addition to using CAPTURE, we verified the results with an independent in-house analysis script, which confirmed all findings reported here within the expected statistical uncertainties. 

For all datasets, the imaging process was carried out using the CASA task $\tt{tclean}$, while the image visualization was performed with CARTA (Cube Analysis and Rendering Tool for Astronomy, \citealt{carta_1, carta_2}). The images were produced using a $\tt{briggs}$ weighting, with $\tt{robust}=0$. A few rounds of self-calibration were also performed to improve the images; however, we did not generally see any significant improvement on the final background noise, probably due to the absence of signal in the phase centre. Table \ref{tab:obs} shows the background level for Stokes I ($\sigma_I$) and V ($\sigma_V$), obtained in a region around the source position, avoiding bright sources.
In App.~\ref{app:images}, we show the maps made by combining all the observations available for a target at a particular band.

\subsection{Flux statistics and search for bursts}

The image over hours-long observation can in principle detect persistent or slowly varying emission, but it can easily miss short-duration intense ECM bursts, ubiquitous in planets \citep{zarka04} and UCDs (e.g., \citealt{route13}). 
In order to look for such short-term events, the time variability has also been investigated for each observational epoch. We used the CASA task $\tt{visstat}$ that calculates the flux statistics in the visibility plane. We looked for potential variability in the real amplitude for LL and RR correlators (being the imaginary part a proxy of the background noise) by integrating the data over time intervals $\delta_t$ much smaller than the on-source time $T$.

If this preliminary light curve revealed any potentially interesting signal, we performed additional checks to exclude emission from other sources or RFI: we looked at the dynamic spectrum in the visibility plane, to see the range of frequencies of the transient signal, using the CASA task $\tt{specflux}$ and $\tt{imstat}$, and we checked the image plane, restricted to the range of time and frequencies of relevant signals, to pinpoint its spatial origin.

\section{Results}\label{sec:results}

\subsection{Upper limits}\label{sec:ul}

We did not find any sign of radio emission at the positions of the proposed targets. In Table \ref{tab:obs}, we provide the background noise for each target, integrated over each session and combining all the observing epochs, in both Stokes I ($\sigma_{I}$) and V ($\sigma_{V}$), together with the isotropic luminosity upper limits $L_\nu^V=4\pi d^2(3\sigma_V)$ for the combined images. The background levels we achieve in the clean images of each observation were typically of the order $10-40~\mu$Jy/beam. An exception is the case of HD 153557 and 2M0122-2439, where the target field was heavily contaminated by the presence of nearby ($\sim$ arcmin away) bright sources, $\sim30$ mJy and $\sim 70$ mJy (see App.~\ref{app:images}). The detailed time variability inspection, with time bins $60$ s and $90$ s, provided non-detection, with $3\sigma$ upper limits on short burst fluxes about $3\sqrt{T/dt}\sim 30$ larger than the rms listed in Table~\ref{tab:obs}. This basically excludes the presence of intense ($\gtrsim$ mJy/beam) bursts, larger than a minute, during any of the observation.

\subsection{Radio flux estimates}\label{sec:flux_estimates}

Out of the three channels mentioned above (fast rotation, stellar wind-magnetosphere interaction, and interaction with satellites) we quantitatively estimate the contribution from the first two, using existing models, explained below.

\subsubsection{Stellar wind-magnetosphere interaction}\label{sec:fluxes}

Radio power from the five most magnetised solar planets shows that it highly correlates with stellar wind energy flux incident on the obstacle, which is the planetary magnetosphere \citep{desch84,zarka92}. The available budget is represented by the incident kinetic and/or magnetic power, which read, respectively \citep{desch84,zarka01,griessmeier07,zarka07}:

\begin{eqnarray}
&& P_{\rm kin} \propto n_{\rm sw}v_{\rm sw}^{3}R_{\rm s}^{2}\,, \label{eq:pinput_kin} \\
&& P_{\rm mag} \propto v_{\rm sw}B_{\perp}^{2}R_{\rm s}^{2}\,, \label{eq:pinput_mag}
\end{eqnarray}
where $n_{\rm sw}$ and $v_{\rm sw}$ are the stellar wind density and velocity, and $B_{\perp}$ is the interplanetary magnetic field perpendicular to the stellar wind flow, all at the position of the obstacle. Note that we neglect the contribution of the orbital velocity to the effective velocity entering in the power (considered in e.g. \citealt{griessmeier07}), since, at large separation considered here, it is much smaller (typically few km/s) than the wind values (hundreds km/s). In analogy with the solar system, the stellar wind velocity is believed to saturate at a constant value at a separation shorter than the ones considered in this study: $v_{\rm sw}$ is considered constant in space. Instead, both $n_{\rm sw}$ and $B_\perp$ vary with the separation from the host star $a$ as:
\begin{eqnarray}
   n_{\rm sw}=n \left(\frac{a}{{\rm 1 au}}\right)^{-2} \\
   B_\perp = B_{\perp 0}\left(\frac{a}{{\rm 1 au}}\right)^{-1}
\end{eqnarray}
where $B_{\perp 0}$ is set by considering the solar value of radial and azimuthal fields at 1 au as $B_r=2.6$ nT, $B_\phi=2.4$ nT, and assuming a Parker model \citep{parker65}, with radial dependence of the radial and azimuthal magnetic field components $B_r\propto 1/a^2$, $B_\phi\propto a^{-1}$, so that the latter always dominate for the large separations (see eqs. 18-22 in \citealt{griessmeier07} for the details). In case of having an estimate for the stellar magnetic field at the surface $B_\star$, we rescale the values, $B_\perp \propto (B_\star/B_\odot)$, where $B_\odot=1.4$ G is the solar value \citep{griessmeier07}. Therefore, for large $a$, both the kinetic and magnetic incident energy are $\propto a^{-2}$, so that the phenomenological trends in the Solar system do not allow us to distinguish between the kinetic or magnetic power as underlying energy budget \citep{zarka07}.
However, \cite{zarka18} studied the Ganymede-induced aurorae on Jupiter, finding a much better correlation with the incident magnetic power than the kinetic one. The \cite{saur13} SPI model also reinforces the idea that the magnetic model might be more physically motivated. However, it is not clear if this conclusion can be generalized, and we consider the kinetic model as well.

If the age is available, $n$, $v_{\rm sw}$ and the field at the stellar surface $B_\star$ are set as in \cite{griessmeier07}:
\begin{eqnarray}
  && n = n_0\left(1 + \frac{t}{\tau}\right)^{-1.86}  \label{eq:n_age} \\
  && v_{\rm sw} = v_0\left(1 + \frac{t}{\tau}\right)^{-0.43}  \label{eq:v_age}  \\
  && B_\star \propto \left(1 + \frac{t}{\tau}\right)^{-0.7}  \label{eq:b_spin_age}
\end{eqnarray}
where $\tau=25.6$ Myr, $n_0=1.04\cdot10^{11}$ m$^{-3}$, $v_0=3971$ km/s, and $B_\star=B_ \odot$ at the solar age $t_\odot=4.5$ Gyr (see \citealt{griessmeier07} and references within).

The quantity $R_{s}$ in eqs. (\ref{eq:pinput_kin}), (\ref{eq:pinput_mag}) is the magnetospheric radius (or stand-off distance), which scales with the magnetic moment $\mathcal{M} \propto B R^3$ as follows: \begin{equation}
{R}_{\rm s} \propto \left( \frac{{\mathcal{M}}^{2}}{{n}_{\rm sw}(a) {v}_{\rm sw} ^{2}} \right)^{1/6},
\label{eq:standoff_distance}
\end{equation}
normalized to the Jupiter values, $R_{s,J}=40~R_{\rm J}$, ${\cal M}_{\rm J}=1.56\times10^{27}$~A~m$^2$, as in \cite{griessmeier07}.

The \cite{stevens05} version of the kinetic model considers the stellar mass loss rate, rewriting $P_{\rm kin}$ in terms of the mass loss rate $\dot{M}$, so that the power is proportional to:
\begin{equation}
    P_{\rm mlr} \propto \dot{M}^{2/3} v_{\rm sw}^{5/3} \mathcal{M}^{2/3} a^{-4/3},
    \label{eq:stevens}
\end{equation}
where the unknown parameters are then the stellar mass loss rate $\dot{M}$ and $v_{\rm sw}$. 
The former may vary from case to case, and across spectral types (see, e.g. \citealt{wood21} for M dwarfs). A proxy to the mass loss rate is the X-ray luminosity $L_x$, following \cite{wood02,stevens05}:
\begin{equation}
    \frac{\dot M}{\dot M_\odot} = \left(\frac{R_\star}{R_\odot}\right)^2\left(\frac{F_{x}}{F_{x,\odot}}\right)^{1.15}~,
\end{equation}
where $R_\star$, $R_\odot=6.957\cdot 10^5$ km are the stellar and solar radii, $F_x=L_x/(4\pi R_\star^2)$ is
the X-ray surface flux (for the Sun, $F_{x,\odot}=3.1 \cdot 10^4$~erg~cm$^{-2}$~s$^{-1}$), and the reference solar mass loss rate is taken as $\dot{M}_{\odot} = 2\cdot10^{-14} M_{\odot}$ yr$^{-1}$. The stellar radius, when not available, is retrieved using the relation with the stellar mass $M_\star$, valid for most main-sequence stars, $R_\star \sim M_\star^{0.8}$.

We have searched for the selected targets in different X-ray databases: the second Rosat All-sky Survey point source catalogue (2RXS, \citealt{boller16}), the eROSITA DR1 (half-sky survey, \citealt{merloni24}), the fourth XMM-Newton serendipitous source catalogue (4XMM, \citealt{webb20}), the Chandra ACIS point-source X-ray catalogue \citep{wang16,wang20}, and the old NEXXUS-2 database for X-ray/UV emitting stars within 25 pc\footnote{\url{https://hsweb.hs.uni-hamburg.de/projects/nexxus/nexxus.html}}. We have found values for half of our selected systems, as detailed below. For the other non-detected sources, the nominal sensitivity of 2RXS and eROSITA DR1 ($\sigma_x\sim 10^{-13}$ and $\sigma_x\sim5\cdot10^{-14}$ erg~cm$^{-2}$~s$^{-1}$, respectively) translates into little informative 3$\sigma$ upper limits of $L_x \sim 3-6\cdot 10^{28}$ erg/s ($\gtrsim 1$ order of magnitude larger than the Sun), for the typical $d=40$ pc sources considered here.

In all cases, when no educated guesses from age or $L_x$ are available, we consider solar-like values for $\dot M$, $n$, $v_{\rm sw}$ and $B_\star$ (eqs.~\ref{eq:n_age}-\ref{eq:b_spin_age}). Note that for our targets there are no magnetic field measurements available, and most of them are likely slow rotators (except 2M0122-2439, for which we use a higher value of $B_\star$). In any case, given the large uncertainties, we explore about one order of magnitude around the fiducial values of $n_{\rm sw}(a)$, $v_{\rm sw}$, $B_\star(a)$, $\dot M$, which are the free parameters of the models.

\subsubsection{Scaling with brown dwarf rotation}

Since rotation certainly plays a role in both planets and isolated UCDs, we also consider the phenomenological scaling law for solar planets by \cite{farrell99} (called model 2), which includes a dependence on the rotation $\omega=2\pi/P_{\rm rot}$ (where $P_{\rm rot}$ is the planet's or brown dwarf's spin period) and fits well the radio power from solar planets:

\begin{equation}
    P_{\rm far} \propto \omega^{0.79}M^{4/3}a^{-1.6}. \label{eq:pfarrell}
\end{equation}
The physical motivation for the inclusion of the rotation was its alleged importance in setting the magnetic moment, which, however, for fast rotators, arguably like the objects considered here, is thought to be independent of rotation (see discussion above). Here, we consider this law as a heuristic proxy model to the combined effect of rotation and interaction with the wind, which could act as source of plasma. Several phenomenological models with different powers of mass, separation and rotation rate have been proposed (see Appendix of \citealt{farrell99}), but here we only consider this one for simplicity.

We also consider the more physically-based model for the  rotation-powered Jupiter auroral oval presented by \cite{hill01} and e.g. applied to $\beta$ Pictoris b by \cite{shiohira24}. It is based on the calculation of the Birkeland (field-aligned) currents for Jupiter, and relies on physical parameters as follows:

\begin{equation}\label{eq:poval}
    P_{\rm oval} \propto (\dot M_{\rm bd}\Sigma_p)^{1/2}\omega^2 R^3 B,
\end{equation}
where the additional uncertain parameters (besides $R$, $\omega$, and $B$) are the mass loss of the fast-rotating object $\dot M_{\rm bd}$ (not be confused with the stellar mass loss), and the height-integrated Pederson conductivity $\Sigma_p$, which depends on the ionospheric ionization, i.e. on the X-ray/ultraviolet flux received. The reference values for Jupiter are $\dot M_{\rm J}=2000$ kg~s$^{-1}$, $\Sigma_{\rm J}=0.6$ S. Due to the general lack of constraints on them, we shall treat their product $(\Sigma_p\dot M_{\rm bd})$ as an effective free parameter, together with $\omega$.


\subsubsection{Flux normalization}

Only a fraction of the incident power is dissipated and converted into radio emission, defined by a conversion efficiency $\eta_{\rm kin}$ or $\eta_{\rm mag}$. In the absence of direct constraints, these efficiencies are scaled with the Jovian case. Observations of Jupiter indicate values ranging from $\sim 10^{-6}$ to $10^{-5}$, depending on the emission regime \citep{zarka98,zarka01,zarka07,griessmeier07, kaya25}, and there is a high uncertainty when extrapolating to other systems.

In this work, the normalization is done directly on the average expected fluxes, like in \cite{stevens05}. Here we conservatively take $\Phi_{\rm J}^{1au}=2.5 \times 10^7$ Jy as the flux at a distance $d=1$ au, representing the peak of the average Jovian spectra as reported by several works \citep{zarka92,farrell99}, and compatible with the Jovian high-activity reference power $P_{\rm J}\sim 2.1\cdot 10^{11}$ W, assumed by e.g. \cite{griessmeier07}, and a factor $\sim 4$ lower than the more optimistic values assumed in \cite{stevens05}:
\begin{equation}\label{eq:flux1}
    \Phi_{\rm i} = \frac{P_{\rm i}}{P_{\rm J}}\Phi_{\rm J}^{1 {\rm au}}\left(\frac{{\rm 1 au}}{d}\right)^2,
\end{equation}
where $\frac{P_{\rm i}}{P_{\rm J}}$ is given by the scaling relations above for each model $i$ above, eqs.~(\ref{eq:pinput_kin}), (\ref{eq:pinput_mag}), (\ref{eq:stevens}), (\ref{eq:pfarrell}),  (\ref{eq:poval}). 
Note that this normalization is different from the exoplanetary predictions by e.g. \citep{farrell99,griessmeier07,ashtari22,shiohira24}, who normalize the power $P_{\rm i}$ to the Jovian value $P_{\rm J}$, and then divide it by $f_b d^2 \Delta \nu$, where $f_b$ is the solid angle of the beamed emission assumed constant (with reference values $f_b=1.6$ sr, valid for the auroral ovals of Jupiter, \citealt{zarka04}, while much lower values are seen for S-bursts, \citealt{queinnec01}), and $\Delta \nu \sim 0.5-1 ~\nu_c$ is bandwidth of the emission, known to be of the order of the cyclotron frequency. In those works, since $\Delta \nu \propto \nu_c \propto B$ and $f_b$ is kept fixed, and the input power of the kinetic and magnetic models scale as $B^{2/3}$, the predicted fluxes would scale as $B^{-1/3}$, which provide unrealistically high flux values for very small fields. We have indeed checked that the phenomenological scaling laws with the flux normalization above, eq. (\ref{eq:flux1}), fits better the average radio flux values of solar planets (reported by e.g. \citealt{zarka92,farrell99}), compared to $P_{\rm i}/(f_b \Delta\nu(B))$. The reason for this discrepancy is probably the assumption of a universal value for $f_b$, which is instead seen to vary, being wider in less magnetized planets \citep{desch84}, thus compensating the $\Delta\nu\propto B$ proportionality. Considering this physical effect likely makes the normalization $P_{\rm i}/(f_b \Delta\nu)$ qualitatively similar to the direct use of $P_i$ (scaled by eq. \ref{eq:flux1}), as we do here.

The widespread use of a $B$-independent, constant value of $f_b$ is at the basis of the counter-intuitive (and physically questionable) results of flux decreasing with increasing $B$, found in many past works, see e.g. the discussion at the end of Sect. 2 in \cite{shiohira24}, when comparing their estimates with \cite{ashtari22}.

\begin{figure}
\centering
\begin{subfigure}{0.45\textwidth}
\centering
\includegraphics[width=.85\linewidth]{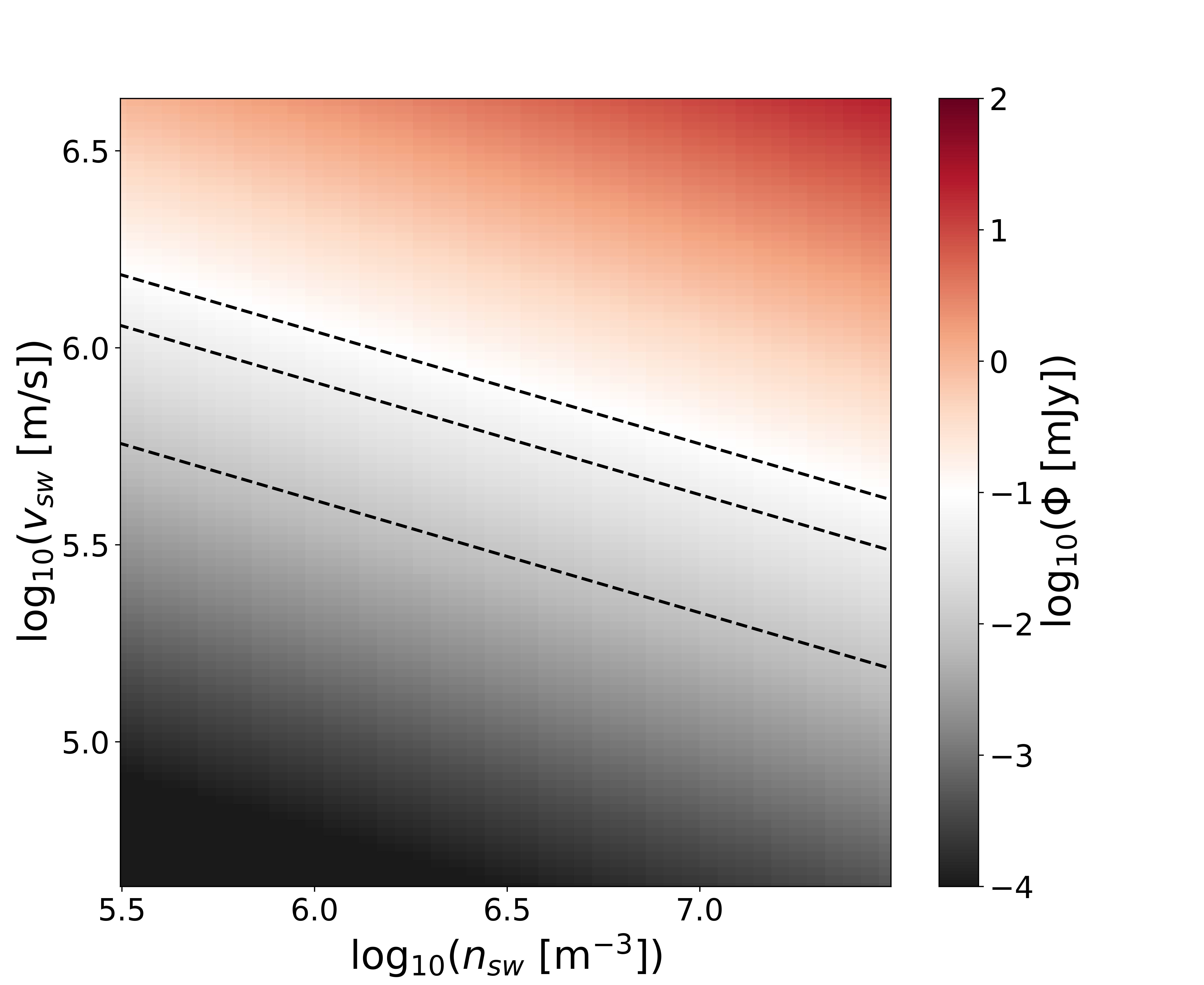}
\subcaption{Kinetic model}
\label{fig:flux_estimates_kinetic}
\end{subfigure}
\hfill
\begin{subfigure}{0.45\textwidth}
\centering
\includegraphics[width=.85\linewidth]{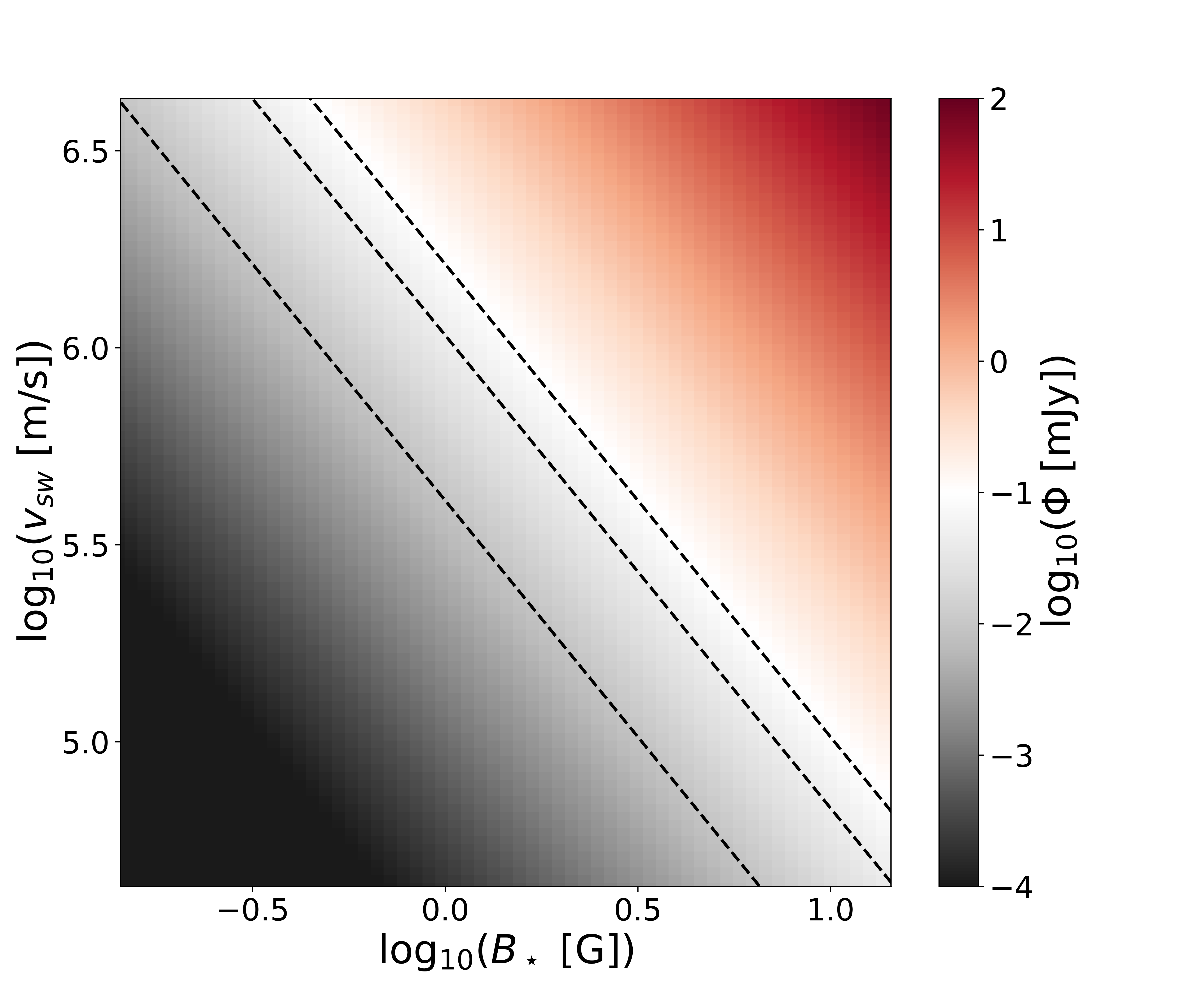}
\subcaption[]{Magnetic model}
\label{fig:flux_estimates_magnetic}
\end{subfigure}
\hfill
\begin{subfigure}{0.45\textwidth}
\centering
\includegraphics[width=.85\linewidth]{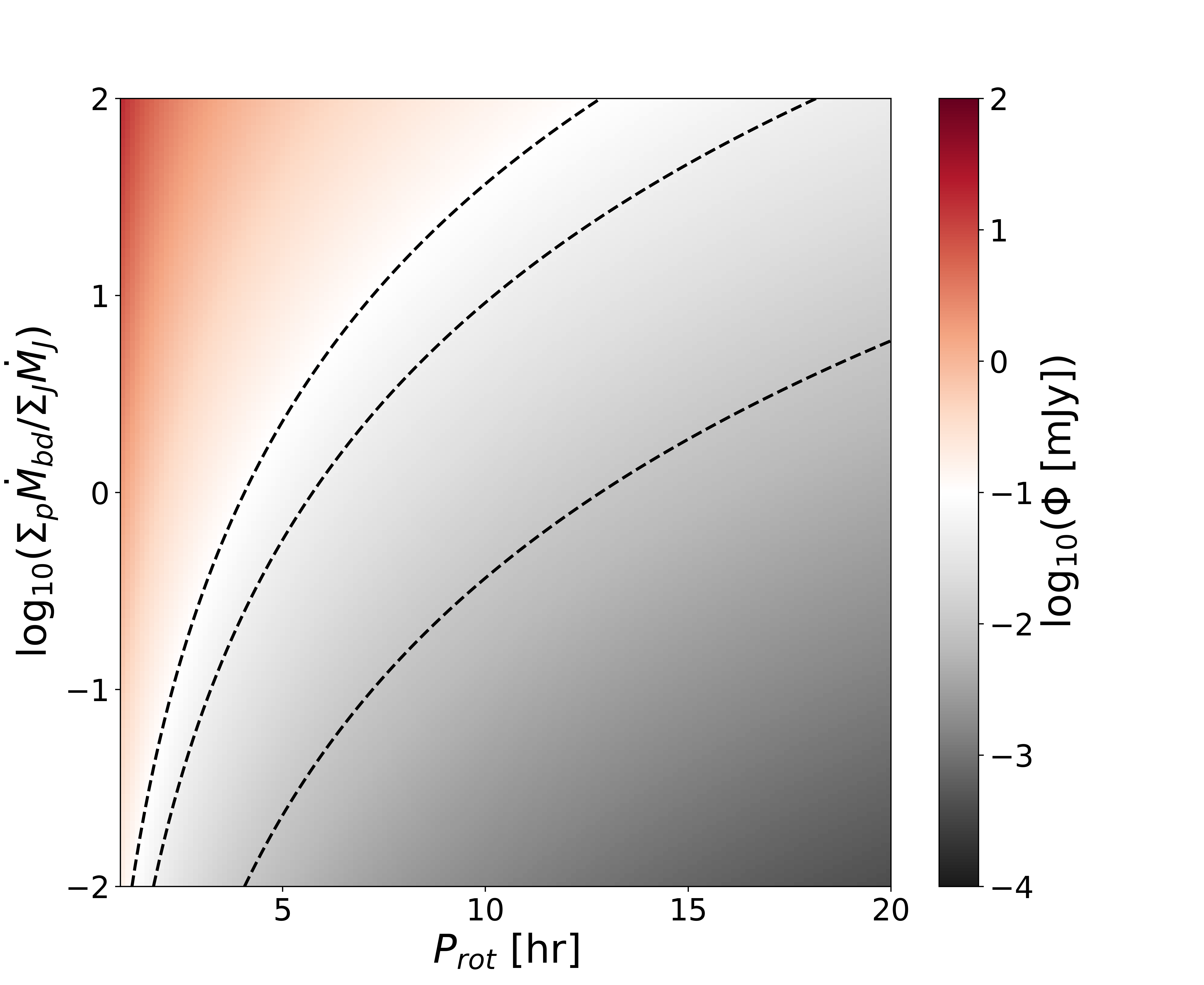}
\subcaption{Rotation-powered auroral model}
\label{fig:flux_estimates_oval}
\end{subfigure}
\caption{Flux estimates of BD-00 4475 B for the different physically based models, from top to bottom: as a function of wind velocity $v_{\rm sw}$ and density $n_{\rm sw}$ for the the kinetic model, wind velocity and stellar surface magnetic field for the magnetic model, and $P_{\rm rot}$ and $\Sigma_p\dot M_{\rm bd}$ for the rotation-powered model. The black dashed lines in the plots correspond to 10, 50 and 100~$\mu$Jy.}
\label{fig:flux_estimates}
\end{figure}

\subsection{Comparison with models}

\begin{table*}
\centering       
\begin{tabular}[h]{cccccccccccccccccc}
\hline
Target & $a$ & $\nu_{c}^p$ &  $B_\star$ & $\log n_{\rm sw}$ & $v_{\rm sw}$ & $R_s$ & $R_\star$ & $\log L_x$ & $\log{\dot M}$ & $\phi_{\rm far}$ & $\phi_{\rm oval}$ & $\phi_{\rm mlr}$ & $\phi_{\rm kin}$ & $\phi_{\rm mag}$ \\
name & [au] & [GHz] & [G] & [m$^{-3}$] & [km/s] & [$R_{\rm J}$] & [$R_\odot$] & [erg/s] & [$M_\odot$/yr] & [$\mu$Jy] & [$\mu$Jy] & [$\mu$Jy] & [$\mu$Jy] & [$\mu$Jy] \\
\hline
HD 153557 d &  21.4 &  0.66 &     {\em 1.4} &   4.2 &   {\em 429} &   193 & 0.8 & 28.40 & -12.38 &  16 & 50 &    19&    2.5 &  2.5\\
2M0122-2439 B$^\star$ &  52.0 &  0.59 &    16 &   6.2 &  1881 &    53 & 0.4 & 28.80 & -11.84 &  1.4 &    35 & 41&  449 &     4.8\\
HD 26161 b &  23.2 &  1.41 &     {\em 1.4} &   4.1 &  {\em 429} &   256 & 1.4 &  - & {\em -13.70} &   7.2 &  21 &  0.7&    0.7 &     0.7 \\
BD-00 4475 B &   1.5 &  1.23 &  {\em 1.4} &   6.5 &   {\em 429} &    97 & 0.8 &  - & {\em -13.70} &    451 &    16 & 24&   24 &    24\\
$\nu$ Oph b &   1.8 &  0.53 &     {\em 1.4} &   6.3 & {\em 429} &    79 & 14.6 &  - & {\em -13.70} &     94 &  6 &   9&    9 &     9\\
$\nu$ Oph c &   5.9 &  0.59 &   {\em 1.4} &   5.3 & {\em 429} &   122 & 14.6 & - & {\em -13.70} &    16 &  7 &     2&    2 &     2\\
GJ 3626 B &   4.0 &  0.83 &     {\em 1.4} &   5.6 &   {\em 429} &   119 & 0.5 & 26.98 & -13.94 &    191 & 39 &   12&   17 &    17 \\
\hline
\end{tabular}

\caption{Fiducial parameters adopted for the systems (see text for references) and representative radio flux estimates. Column indicates: target name, assumed separation at the time of observations (see text), estimated cyclotron frequency at the polar surface of the brown dwarf, magnetic field of the star, density and velocity of the wind, magnetospheric radius, stellar radius, logarithm of the X-ray luminosity, logarithm of the stellar mass loss rate, and the corresponding estimated fluxes for the five models here considered. Values in italics correspond to the solar/Jovian values for $B_\star$, $v_{\rm sw}$, $\dot M$, due to absence of constraints, and are taken from \citealt{griessmeier07} and references within (see text). As a comparison, the Jupiter flux at $d=40$ pc is 0.37\,$\mu$Jy with the normalization here assumed (see text). \\
$\star$ This object is the only one for which we use the reported age to guess the wind density (in logarithm) and velocity, and for which there is a possible measurement of the rotation $P_{\rm rot}=6$ h \citep{zhou19}. For the other targets, a Jovian-like rotation value $P_{\rm rot}=10$ h is assumed.}

\label{tab:parameters}
\end{table*}

\subsubsection{Expected fluxes}

 Table \ref{tab:parameters} reports the parameters adopted for the flux estimation. For eccentric systems with well-constrained orbital solutions, we consider the expected separations of the brown dwarf at the time of the observation; otherwise, we adopt the semi-major axis.  The expected electron cyclotron frequency $\nu_c$ is evaluated considering the magnetic field estimate eq.~(\ref{eq:magnetic_scaling}), as explained in Sect. \ref{sec:freq_mf}. Note that in the two cases where only $M\sin i$ is available (not $M$), the reported value $\nu_c$ is actually a lower limit. In only two cases, we have an indication of the age, which can be used as an educated guess for the stellar wind properties as explained above. We did not find in the literature precise measurements of the rotation period of the stars (which can be a proxy for the expected stellar field and X-ray luminosity, \citealt{reiners14}), except in the only case where the age and $L_X$ is also available. 

Table \ref{tab:parameters} also reports the resulting flux estimations of each target for the considered models, for the set of fiducial values of the free parameters. A few remarks are needed. The magnetospheric radius is always larger than Jupiter's one (40 $R_{\rm J}$), due to the large assumed magnetic fields $B$ and the low ram pressure from the wind. This means that magnetospheric electrodynamics is expected to be qualitatively similar to isolated brown dwarfs, with rotation playing a major role. Note that in cases with no available parameter to guide the guess, the fluxes from the three kinetic/magnetic models are identical, since they scale in the same way with the separation $a$ and the reference values are the solar ones. In the other cases, and for the \cite{farrell99} model, estimates are different, due to the adopted reference parameters. In some cases, notably for one version of the kinetic model of 2M0122-2439 B and the Farell models of BD-00 4475 B and $\nu$ Oph b, interesting values of $\sim 0.1-0.5$ mJy are predicted, with the nominal parameters adopted. In most cases, the values are around ${\cal O}(10)\, \mu$Jy, admittedly at or below the limit of the current detectability. However, uncertainties are large: Table \ref{tab:parameters} shows the fluxes for the fiducial values only. 

The sensitivity of the fluxes on the free parameters are shown in Fig.~\ref{fig:flux_estimates}, for one of the targets, BD-004475 B, as a representative example. We show the physically based models: the kinetic model, the magnetic model, and the \cite{hill01} model. The \cite{farrell99} phenomenological model is not shown, since it only depends on $\omega$ (eq. \ref{eq:pfarrell}) and it's then trivial to scale the reference values $\Phi_{\rm far}$ shown in Table \ref{tab:parameters}. Considering a range of one order of magnitude around the fiducial values of the parameters, the estimated flux can typically vary by 2-3 orders of magnitude. The same is true for the other targets.

Considering the large uncertainties of the model parameters, and the possibilities of having temporary flux enhancement by 1-2 orders of magnitude due to e.g. variations of solar winds or favourable rotational phases, sources could be potentially detectable with a few hours exposure with VLA and uGMRT.

\subsubsection{Tentative constraints on the models}

\begin{figure}
\centering

\begin{subfigure}{0.4\textwidth}
\centering
\includegraphics[width=.85\linewidth]{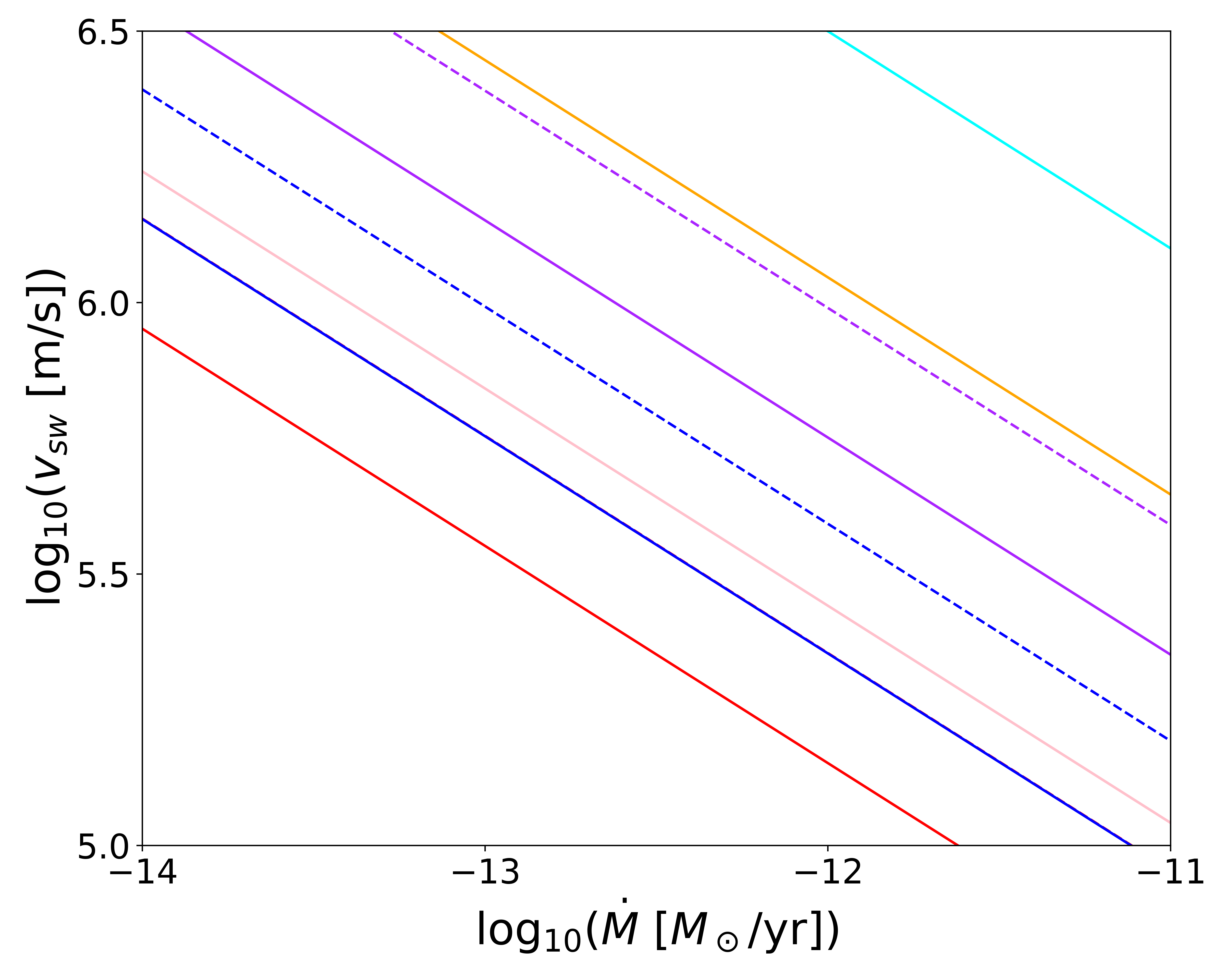}
\label{fig:upperlimits_a}
\end{subfigure}

\vspace{0.3cm}

\begin{subfigure}{0.4\textwidth}
\centering
\includegraphics[width=.85\linewidth]{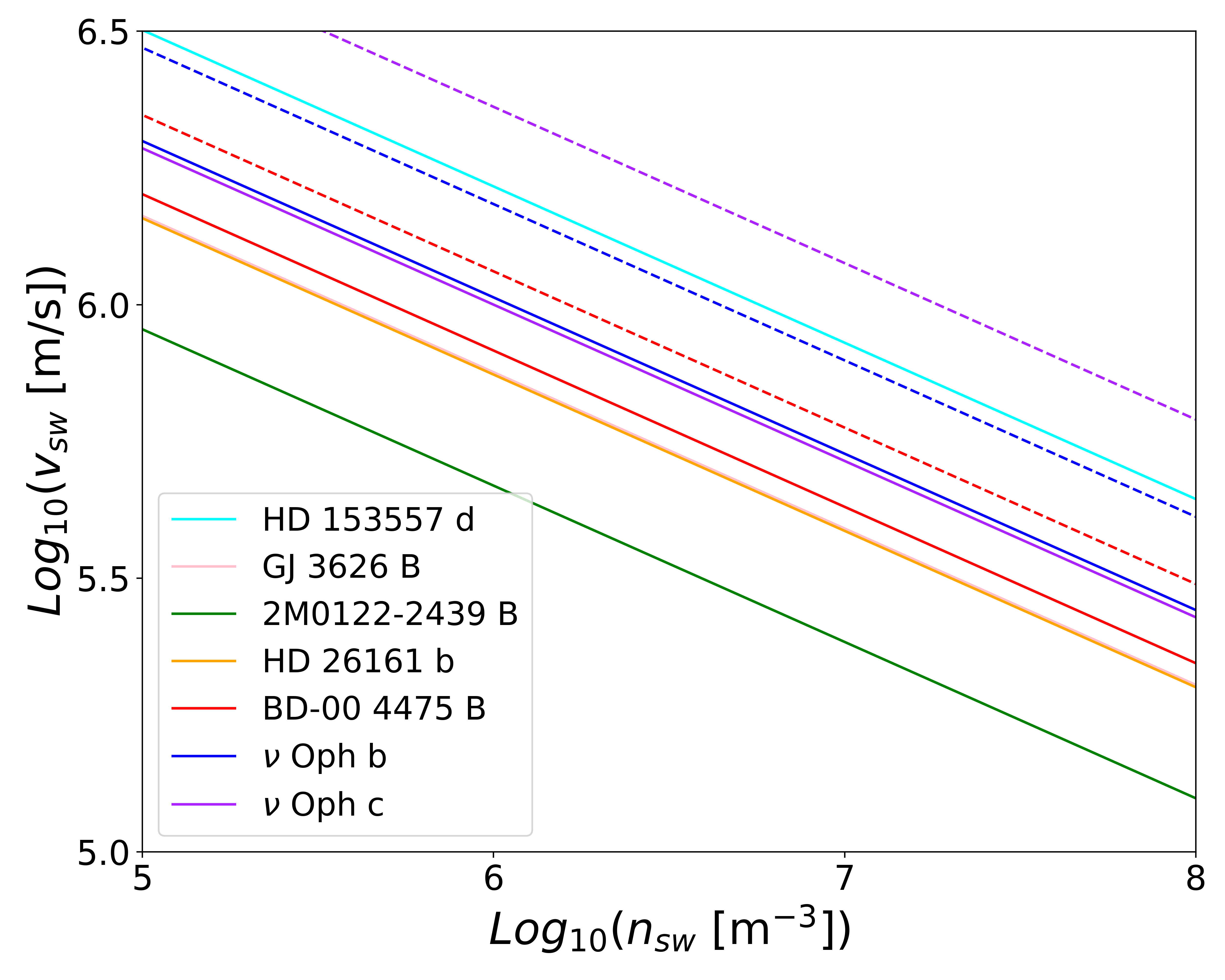}
\label{fig:upperlimits_b}
\end{subfigure}

\vspace{0.3cm}

\begin{subfigure}{0.4\textwidth}
\centering
\includegraphics[width=.85\linewidth]{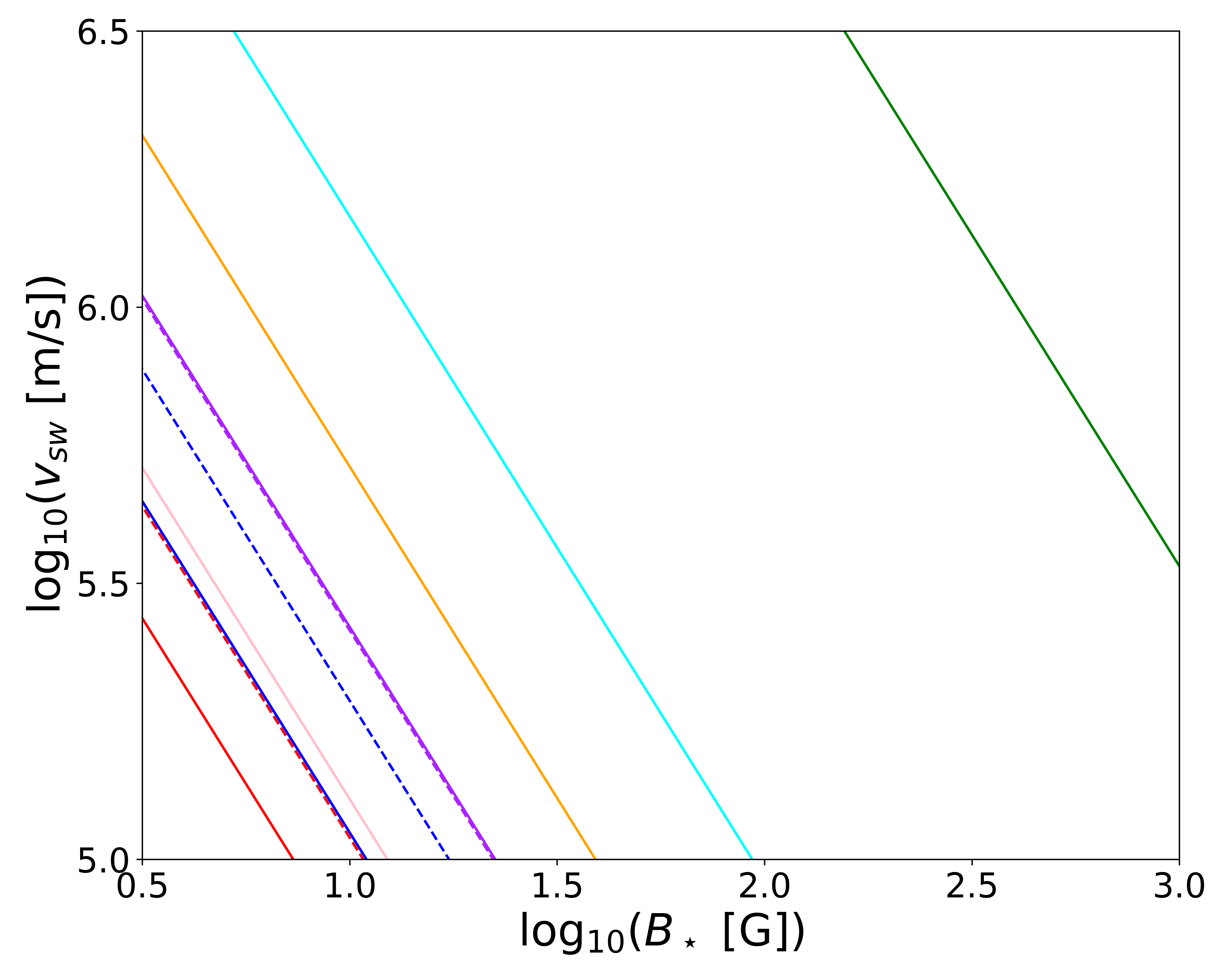}
\label{fig:upperlimits_c}
\end{subfigure}

\vspace{0.3cm}

\begin{subfigure}{0.4\textwidth}
\centering
\includegraphics[width=.85\linewidth]{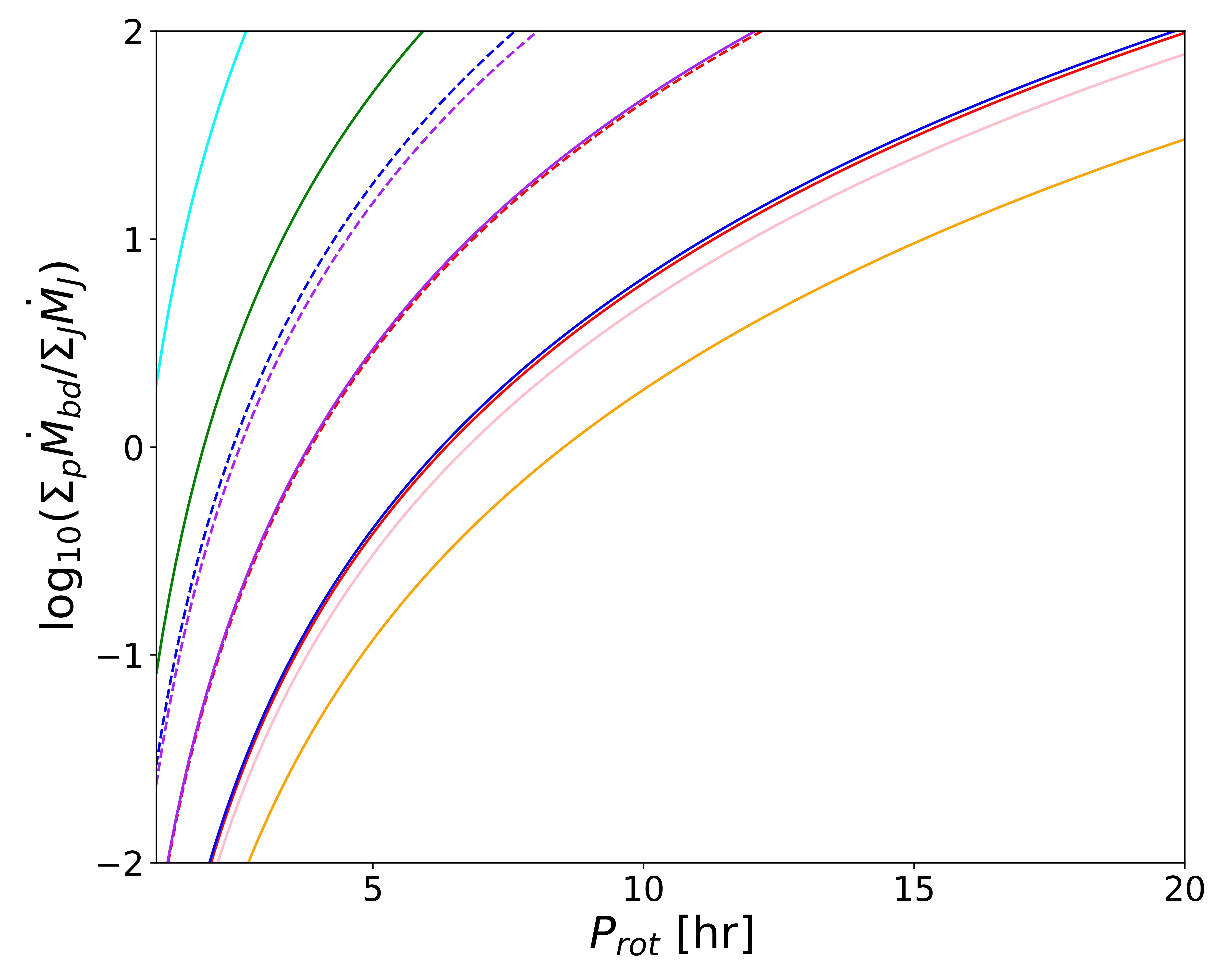}
\label{fig:upperlimits_d}

\end{subfigure}
\caption{Tentative constraints on the parameters of models, from top to bottom: kinetic model as a function of mass loss rate and wind density (eqs. (\ref{eq:pinput_kin}), (\ref{eq:stevens})), magnetic model (eq. \ref{eq:pinput_mag}), and rotation-powered auroral model (eq. \ref{eq:poval}). Lines in the space parameters corresponding to the $3\sigma_I$ upper limits of the combined images for each target (different colours). Assuming that the models are correct, a favourable beaming and no time variability, the regions above the lines are nominally excluded. The dashed lines indicate the observations made with VLA and the solid lines with uGMRT.}
\label{fig:upper_limits}
\end{figure}

For each system, we translate the $3\sigma_{I}$ upper limit into nominal constraints on the plane of the two main free parameters of each of the physical models described above. 
In Fig. \ref{fig:upper_limits} we show the constraints of such parameters for every target (colours, linestyle showing VLA or uGMRT combined observations), where regions above the lines are formally excluded, if one assumes that the model is correct and the lack of emission is not due to intrinsic variability or beaming (see below). 
Note that slightly tighter constraint (curves shifting downwards, and to the left or right depending on the model) could be placed assuming a high circularly polarised fraction $V/I\gtrsim 0.5$ of the expected emission, and using $3\sigma_V$ upper limits in a those cases where strong, unpolarized nearby sources limited the Stokes I sensitivity only (HD 153557 d and 2M0122-2439 B). On the other hand, more loose constraints (curves shifting upwards) would be obtained when considering the higher upper limits by a factor $\sim 30$ on minutes-long bursts (see Sect. \ref{sec:ul}).

These plots indicate that the derived upper limits vary significantly depending on the assumed model and on the target. In some cases, for which the upper limits are pretty high, the constraints are loose (e.g., 2M0122-2439 B in the \cite{stevens05} kinetic model, out of the plotted range, and in the magnetic model). 2M0122-2439 B is also peculiar, compared to the other sources, because it appears to be the most and the least constraining case, depending on the model (cfr. first and second panel). This is due to the wide difference among the expected fluxes for the fiducial parameters of the four models (last four columns of Table \ref{tab:parameters}). The magnetic model (third panel) excludes in most cases values of $B_{\star}\gtrsim 10$ G for Solar-like wind velocity values. This is interesting since, in principle, these results could be compared with observational constraints from Zeeman-Doppler imaging measurements. As far as we know, the only published magnetic field measurements in our targets is for $\nu$ Oph \citep{auriere15,butkovskaya21}, for which there are only upper limits of $B_\star\lesssim {\cal O}(1)$ G. For the rotation-powered auroral model (bottom panel), only fast rotation ($P_{\rm rot}\lesssim 0.5-8$ h, depending on the target) can be excluded, assuming Jovian-like $\Sigma_p\dot M_{\rm bd}$ values. Considering a factor 10 larger value of the latter (e.g., similar $\Sigma_p$ but higher $M_{\rm bd}$ due to $M\gg M_{\rm J}$) would set tighter constraints on the period, $P_{\rm rot}\lesssim 1.5-15$ h.

However, note that these constraints are only tentative, since, as we discuss in the next section, there are several factors which could explain the lack of detection. Therefore, we are not claiming that our observations can really rule out some values of rotation, wind or magnetic properties.

\section{Discussion and conclusions} \label{sec:Discussion}

In this work, we have presented a targeted low-frequency deep-field radio search of $\sim$ 60 h, focused on brown dwarfs with masses in the $\sim 20$–$50~M_{\rm J}$ range, located just above the boundary between giant planets and brown dwarfs. Our selection included nearby ($<50$ pc), so-far unobserved objects with theoretically promising radio flux predictions according to some of the classical models of wind-magnetosphere interaction \citep{griessmeier07,stevens05} and rotation-powered aurorae \citep{hill01}. The motivations for this search were multiple: (i) widely separated brown dwarfs spin fast like in isolation (see solar gas giants), so that both wind-magnetosphere interaction and co-rotation breakdown are potential engines to power radio emission (similar efforts, with very massive long-period planets, are very limited \citealt{cendes22,shiohira24}); (ii) using simple scaling laws, magnetic fields of a few hundred gauss are expected for brown dwarfs, so that the ECM fundamental frequency falls within the 0.3-2 GHz explored here, and the expected radio fluxes (due to either mechanism above) are also larger than for e.g. giant planets; (iii) UCDs have been rarely observed at $\lesssim 1$ GHz, being most studies led at several GHz, i.e. sensitive to much larger fields only; (iv) a detection would bridge the gap in radio characterization between isolated fast-rotating UCDs and solar planets, with exoplanetary radio emission still being elusive.

Despite the optimistic flux estimates made theoretically, our observations yielded only upper limits in both Stokes I and V, with no evidence of persistent emission or transient short ($\gtrsim$ minutes) bursts. The typical sensitivity achieved in our observations for the combined images ($\sigma\lesssim 10\,\mu$Jy/beam in Stokes V for many cases), is consistent with those achieved in previous campaigns. The corresponding upper limits of radio luminosities for the average emission are $L_\nu^V\sim 0.2-10\times 10^{13}$~erg~s$^{-1}$~Hz$^{-1}$, in the same range as the typical detections and upper limits of UCDs \citep{pineda23}, thus not discarding lower flux which has been detected (at a few GHz) in several cases \citep{rose23}. Unfortunately, the arguably most promising candidate within our sample, the direct imaged 2M0122-2439 B, is also the one with the highest rms, due to a nearby background source. Notably, we also exclude short ($\gtrsim$ minutes-long) bursts above $\gtrsim$ mJy level, for all sources.

As discussed in most of similar studies (e.g., \citealt{narang24} and references within) and in the review by \cite{callingham24}, a non-detection of expected ECM coherent emission may result from several reasons. One major factor potentially contributing could be the nature of the emission mechanism itself. ECM, the most likely coherent processes expected to power planetary auroral radio emission, is highly beamed and time-variable. This emission is confined to narrow cones that may not intersect the Earth's line of sight during the time of observations (\citealt{hallinan07}). For instance, Jupiter’s intense decametric radiation, is confined to a $\sim$1.6 sr solid angle (only about 10–13\% of the sky). Therefore, the auroral beam would be observable from Earth only in a small fraction of orientations. This probably explains the derived rate of $\lesssim 10\%$ occurrence of radio-loud (isolated) UCDs. Brown dwarfs orbiting around other stars, targeted in this study, should have a similar occurrence rate, since the the magnetospheric radius is large and the rotation is expected to be rapid, as hinted by the only target in our sample with a 6 h rotation measurement claim \citep{zhou19}. The possible interaction with a stellar wind can give a non-negligible extra energy budget in some cases. Assuming a $10\%$ occurrence rate, not finding emission from 8 objects is not surprising ($0.9^8=45\%$), and at least 22 systems should be observed to secure at least one detection with $> 90\%$ probability, as already derived by \cite{cendes22} in the context of directly imaged, massive planets. This implies that a much larger sample then ours should be targeted before reaching any conclusion on the intrinsic radio brightness of these targets.

There are other important limiting factors. Even if beaming was favourable, ECM signals are often sporadic, occurring in bursts that might be missed during limited-time observations. We argue that this is a less problematic factor, because UCDs can often have both quiescent, partially polarized emission, and rotationally-modulated transient emission. The typical rotation of isolated UCDs, and arguably of our targets, is of some hours, so that our observations, with total on-source time of $\sim 2-12$ h per target, are likely have covered a large fraction of the rotational phase. Note that the tiny coverage of the orbital phase is probably not worrying, since the wind-magnetosphere models do not depend on the orbital phase (except for eccentric systems via separation), and rotation is expected to play a major role. This is at contrast with sub-Alfv\'enic SPI (obviously excluded in our targets due to their large separations from the host stars), for which the planetary rotation is long (days), being tidally locked to the orbit. We can also likely discard free-free absorption as responsible for the non-detections in our case. While dense stellar wind environments may absorb low-frequency radio waves before they escape the system (see \citealt{pena25} in the context of short-period planets), the separations of the targets here presented are such that the density is not high enough to have $\nu_p\gtrsim \nu_c$, unless considerable amount of plasma is provided by e.g. satellites.


A more critical consideration is the degree of uncertainty in the expected frequency (here derived by a simple magnetic-mass scaling law), and in the model parameters that have been used to predict the contribution of wind-magnetosphere interaction to radio fluxes. Theoretical estimates of the latter depend heavily on quantities such as the stellar wind velocity and density (or the mass loss rate), which are often poorly known. We have tentatively constrained the maximum values of such parameters corresponding to a quiescent emission, i.e. using the upper limit of integrated images. However, we need to keep in mind that the wind properties in the Sun vary in time \citep{johnstone15}, space (see e.g. the Parker wind model \citealt{parker65}), and age (which is often poorly constrained), and that, in general, there are several simplifications in the models, starting from the normalization of values based essentially on the Jovian case and the solar planets. The combination of such parameter uncertainties can propagate into orders-of-magnitude errors in the predicted radio flux (Fig. \ref{fig:flux_estimates}).

In conclusion, our findings add to the growing list of non-detections of sub-stellar objects in radio, being compatible with a $\lesssim 10\%$ occurrence rate seen in UCDs. This underscores the challenges of detecting ECM auroral signals from exoplanetary or brown dwarf companions. The flux estimates for our targets are at the limit or below the current detectability for a wide range of the model parameters, but arguably within the SKAO sensitivity. Future progress may rely on a combination of refined modelling, coordinated multi-epoch observations with a more systematic search of promising targets, and large-scale statistical searches through cross-matched datasets. In this sense, the recent and upcoming surveys and campaigns by LOFAR \citep{callingham23,yiu24} and ASKAP \citep{pritchard21,driessen24}, and the advent of SKAO have the potential of having a much more systematic and deep monitoring of wide-orbiting brown dwarfs and massive planets. Directly imaged ones can offer the advantage of possibly having hints of their rotation rate \citep{zhou19}.

In any case, as highlighted by tens of previous exoplanetary radio searches (e.g., \citealt{callingham24} and references within), any radio detection from this kind of system would imply the need of a follow-up to pinpoint the stellar versus companion origin, especially in the case of chromospherically active stars. In our cases, the orbital modulation (which is searched for in short-orbit exoplanetary systems, e.g. \citealt{perez21,turner24}) cannot be proved due to the tiny coverage of the long orbits. 
However, a detectable smoking gun would be a clear $\sim$ hours modulation of the signal not ascribable to the measurable, slower stellar rotation: it would be a hint for the emission being powered by fast rotation of the companions, as in isolated UCDs.

\section*{acknowledgements}
RPM, SK, DV, OM, JMG, ASM and FdS's work has been supported by the program Unidad de Excelencia María de Maeztu, awarded to the Institut de Ciències de l'Espai (CEX2020-001058-M). RPM and SK carried out this work within the framework of the doctoral program in Physics of the Universitat Aut\`onoma de Barcelona. RPM, SK, OM, and DV are supported by the European Research Council (ERC) under the European Union’s Horizon 2020 research and innovation programme (ERC Starting Grant "IMAGINE" No. 948582, PI: DV). ASM acknowledges support from the RyC2021-032892-I grant funded by MCIN/AEI/10.13039/501100011033 and by the European Union `Next GenerationEU’/PRTR. OM, JMG, ASM and GB acknowledge support from the PID2023-146675NB-I00 (MCI-AEI-FEDER, UE) program. FDS acknowledges support from a Marie Curie Action of the European Union (Grant agreement 101030103). MD acknowledges financial support through INAF mini-grant fundamental research funding for the year 2022. KH acknowledges the FED-tWIN research program STELLA (Prf-2021-022), funded by the Belgian Science Policy Office (BELSPO). This research has made use of the NASA Exoplanet Archive, which is operated by the California Institute of Technology, under contract with the National Aeronautics and Space Administration under the Exoplanet Exploration Program. We thank Andreas Quirrenbach and Jose A. Caballero for their useful comments, and the staff of the uGMRT that made these observations possible. uGMRT is run by the National Centre for Radio Astrophysics of the Tata Institute of Fundamental Research, India. 

\section*{Data Availability}
Radio data of the observations here analysed are publicly available from GMRT and VLA databases. Clean images are available upon reasonable request to the authors. The script for the flux calculations and production of the related plots here shown is available at {\url{https://github.com/Simranpreet-buttar/Radio-flux-estimates}}.

\bibliographystyle{mnras}
\bibliography{radio}

\newpage

\appendix

\section{Deconvolved images}\label{app:images}

\begin{figure*}
\centering
\setlength{\tabcolsep}{6pt} 

\newcommand{\panelH}[3][5.7cm]{
  \begin{tikzpicture}[baseline=(img.north)]
    \node[inner sep=0, outer sep=0] (img)
      {\includegraphics[height=#1, keepaspectratio]{#2}};
    \node[anchor=north, font=\scriptsize, text=blue,
          fill=white, rounded corners=1pt, inner sep=1pt,
          fill opacity=0.6, text opacity=1,
          yshift=-4pt] at (img.north) {#3};
  \end{tikzpicture}%
}

\scalebox{0.99}{
\begin{tabular}{@{}p{0.485\textwidth} p{0.485\textwidth}@{}}
\centering
\panelH[5.7cm]{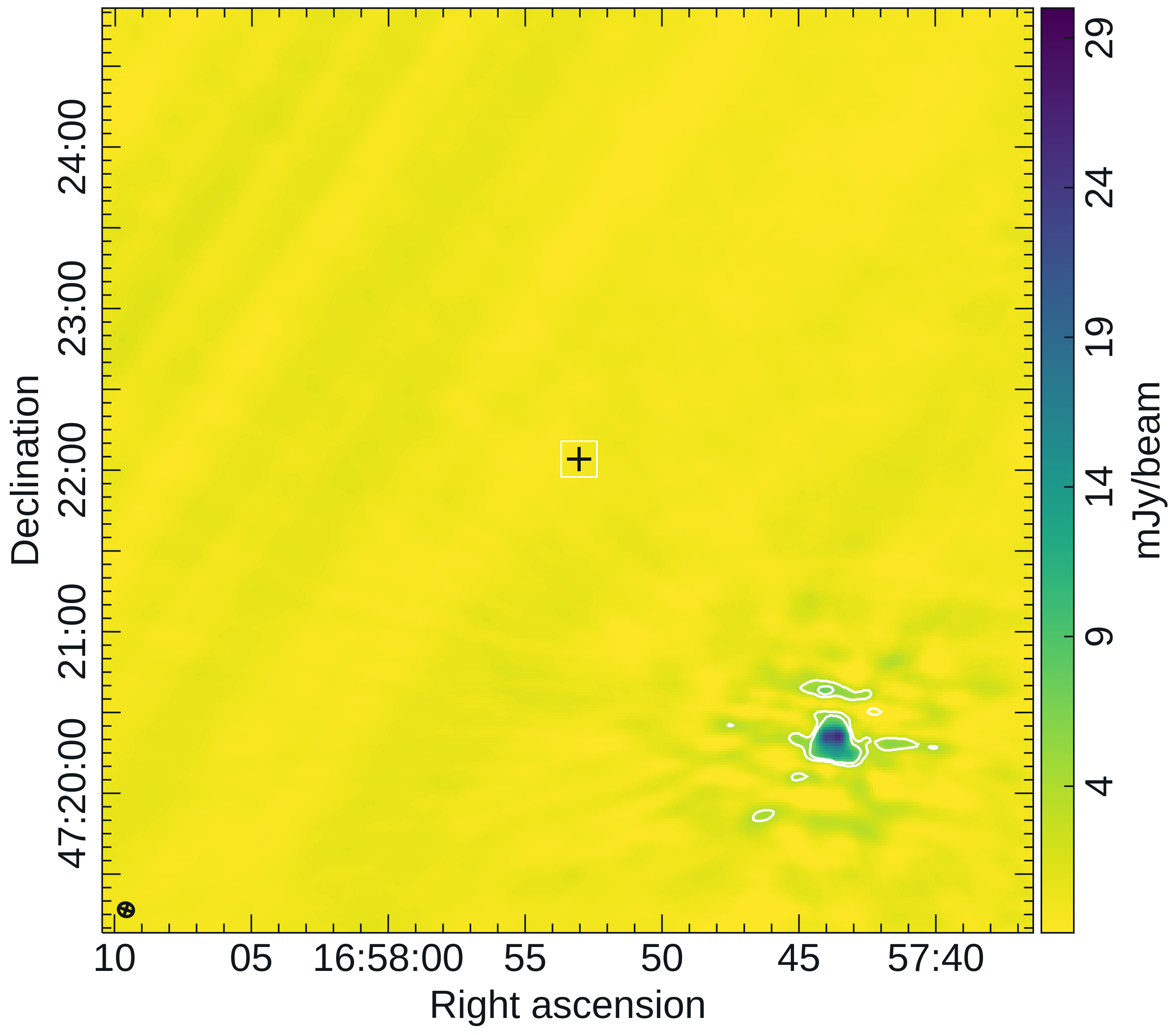}{uGMRT band 4 (0.55-0.75 GHz): HD 153557} &
\panelH[5.7cm]{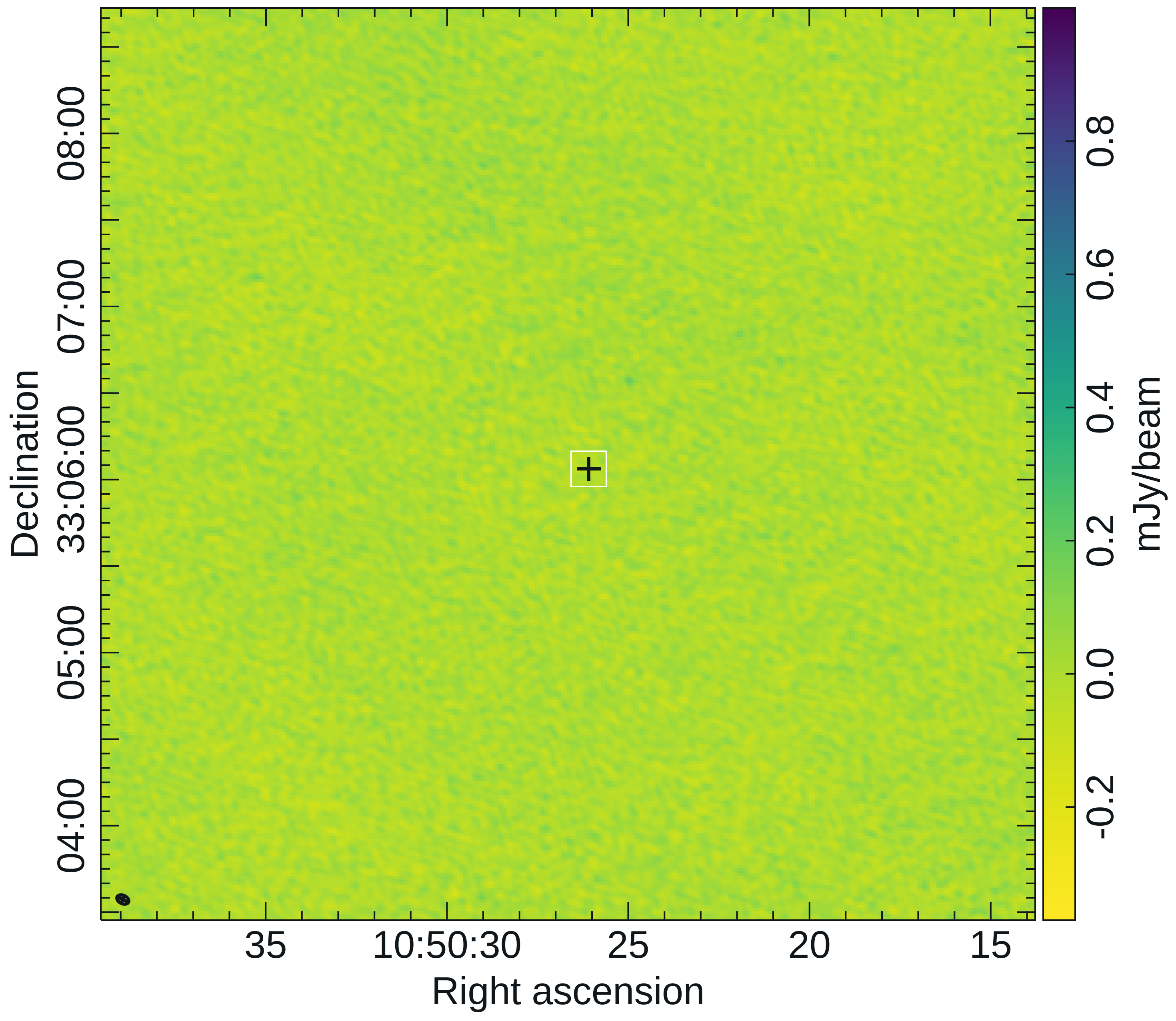}{uGMRT band 4 (0.55-0.75 GHz): GJ 3626} \\
\centering
\panelH[5.7cm]{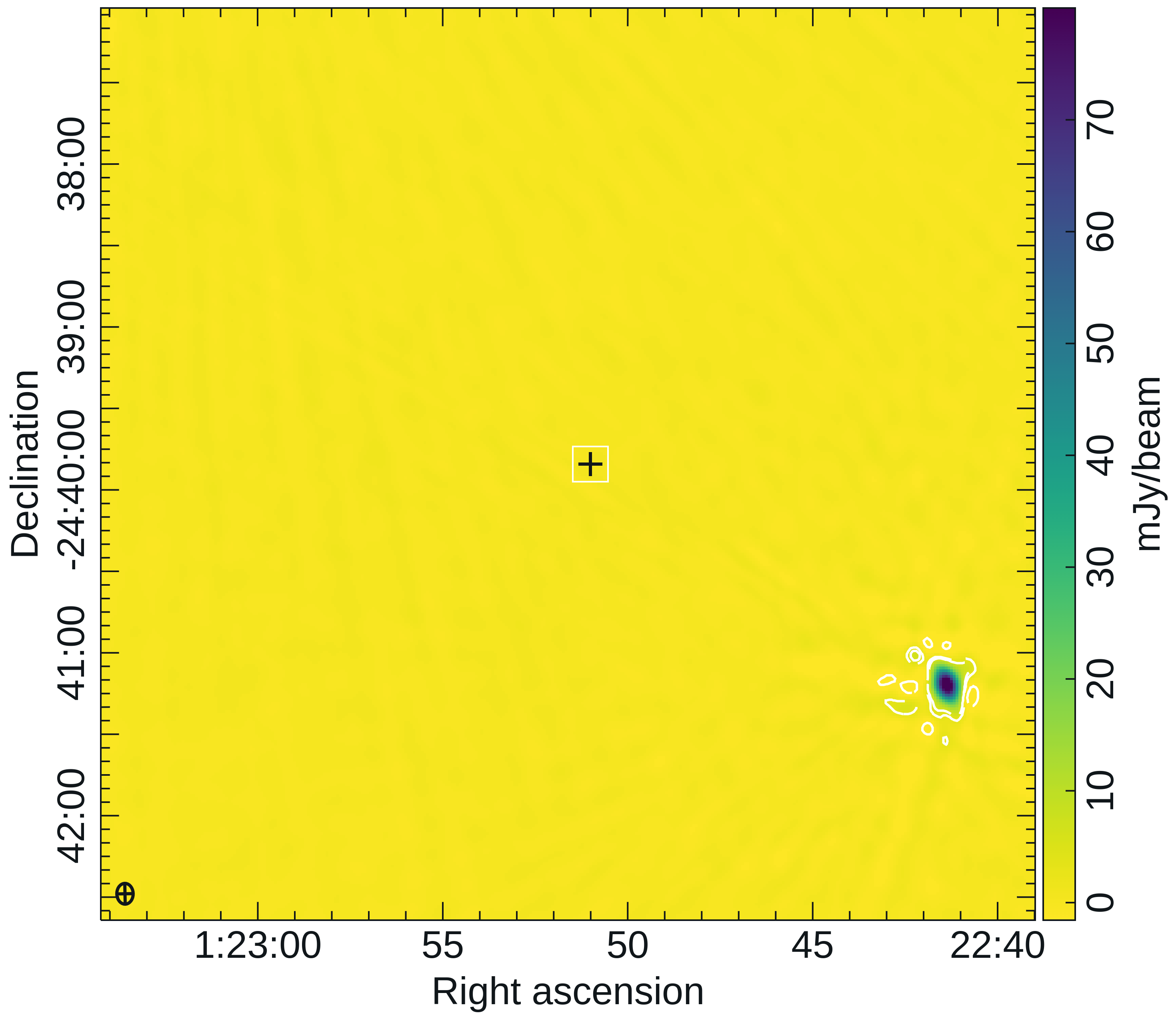}{uGMRT band 3 (0.3-0.5 GHz): 2M0122-2439} &
\panelH[5.7cm]{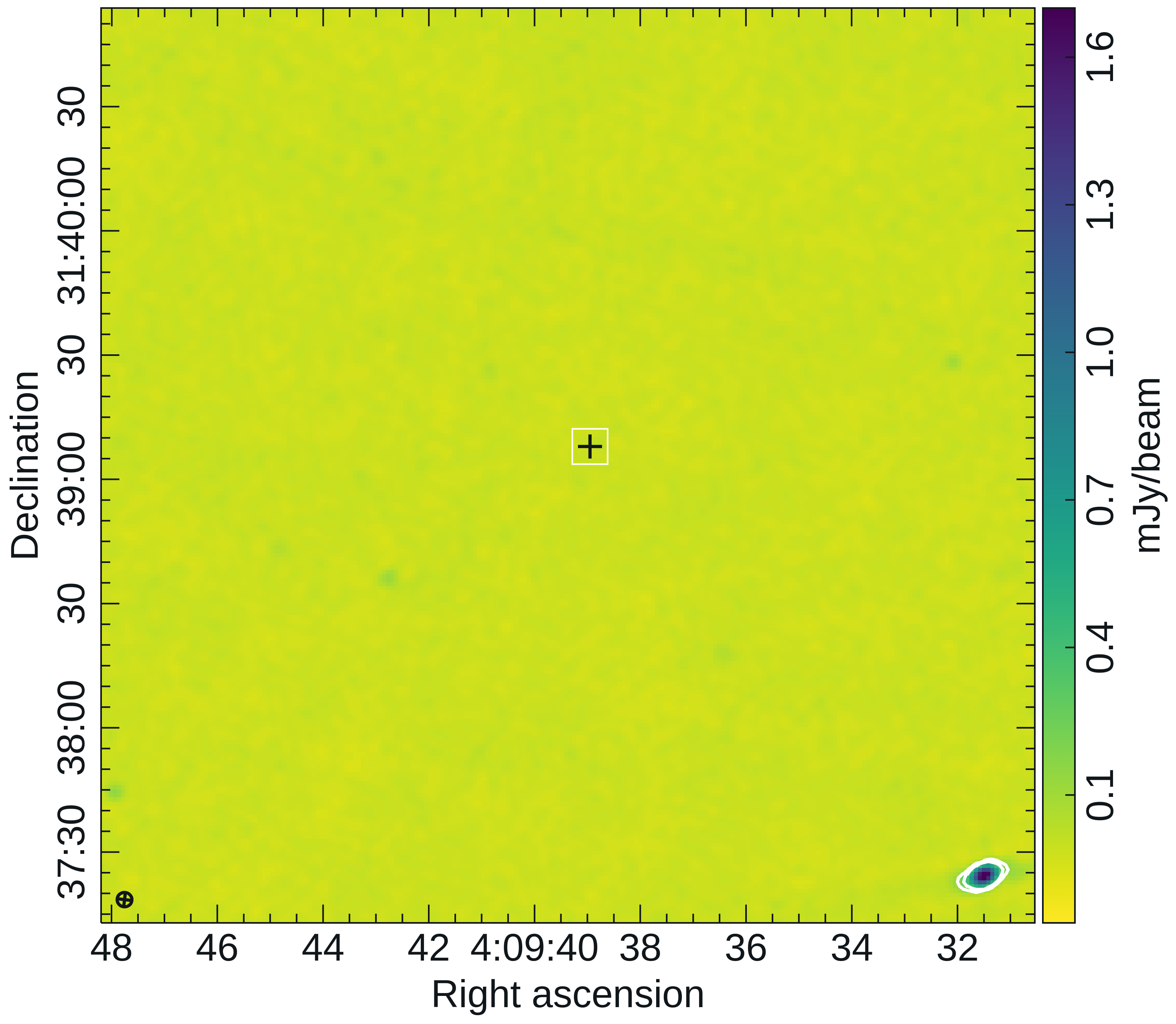}{VLA band L (1-2 GHz): HD 26161} \\
\centering
\panelH[5.7cm]{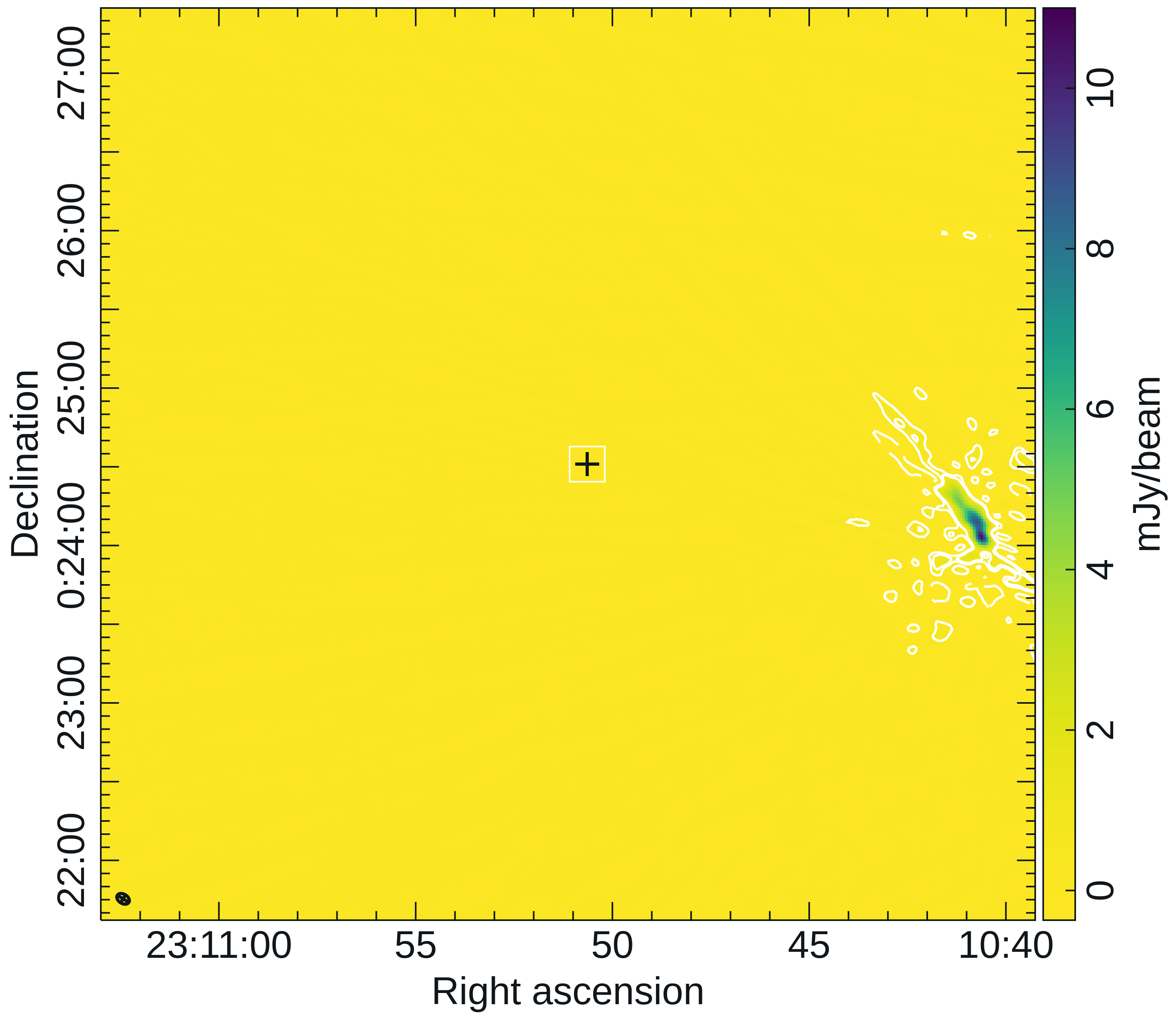}{uGMRT band 4 (0.55-0.75 GHz): BD-004475} &
\panelH[5.7cm]{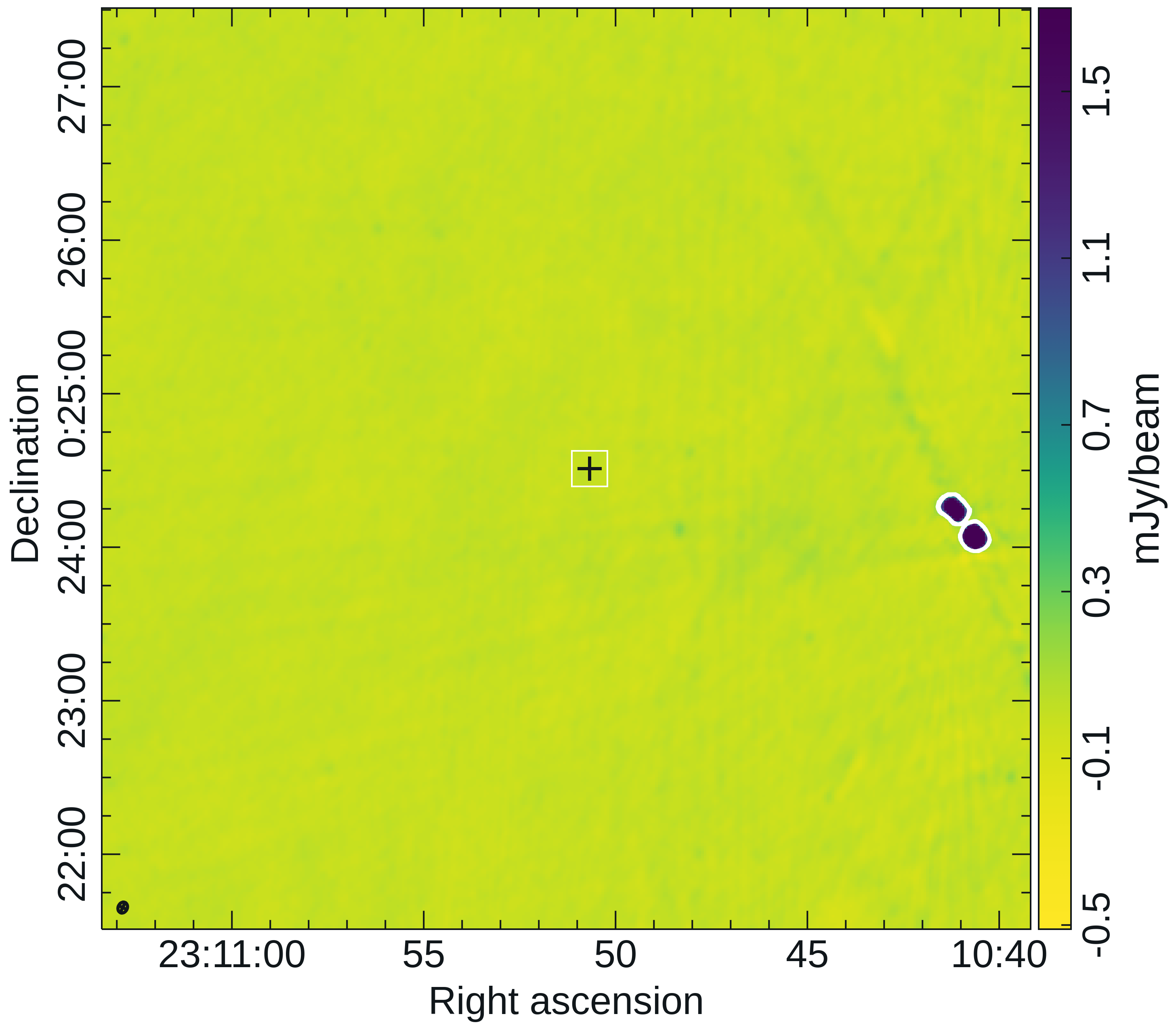}{VLA band L (1-2 GHz): BD-004475} \\
\centering
\panelH[5.7cm]{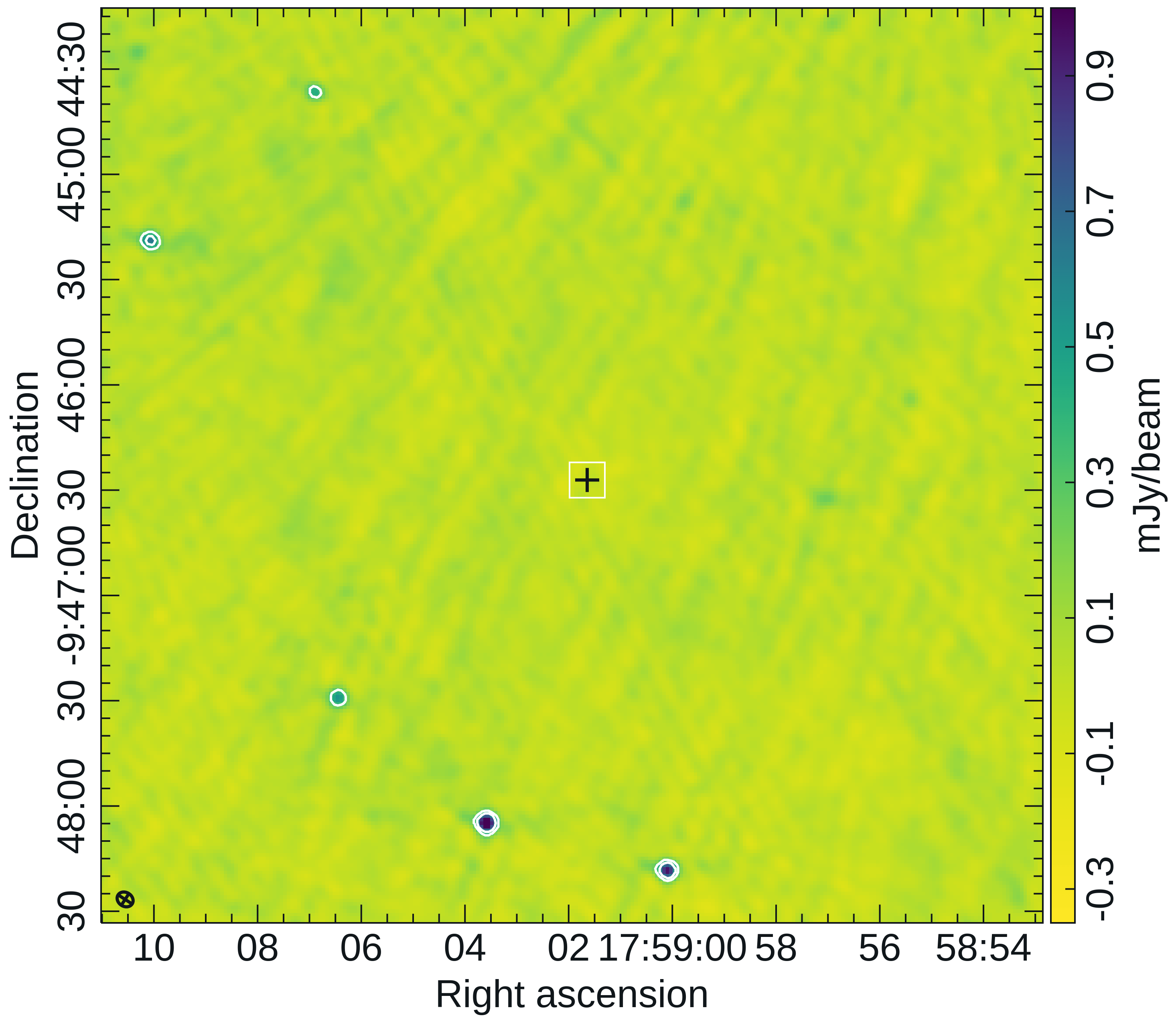}{uGMRT band 4 (0.55-0.75 GHz): $\nu$ Oph} &
\panelH[5.7cm]{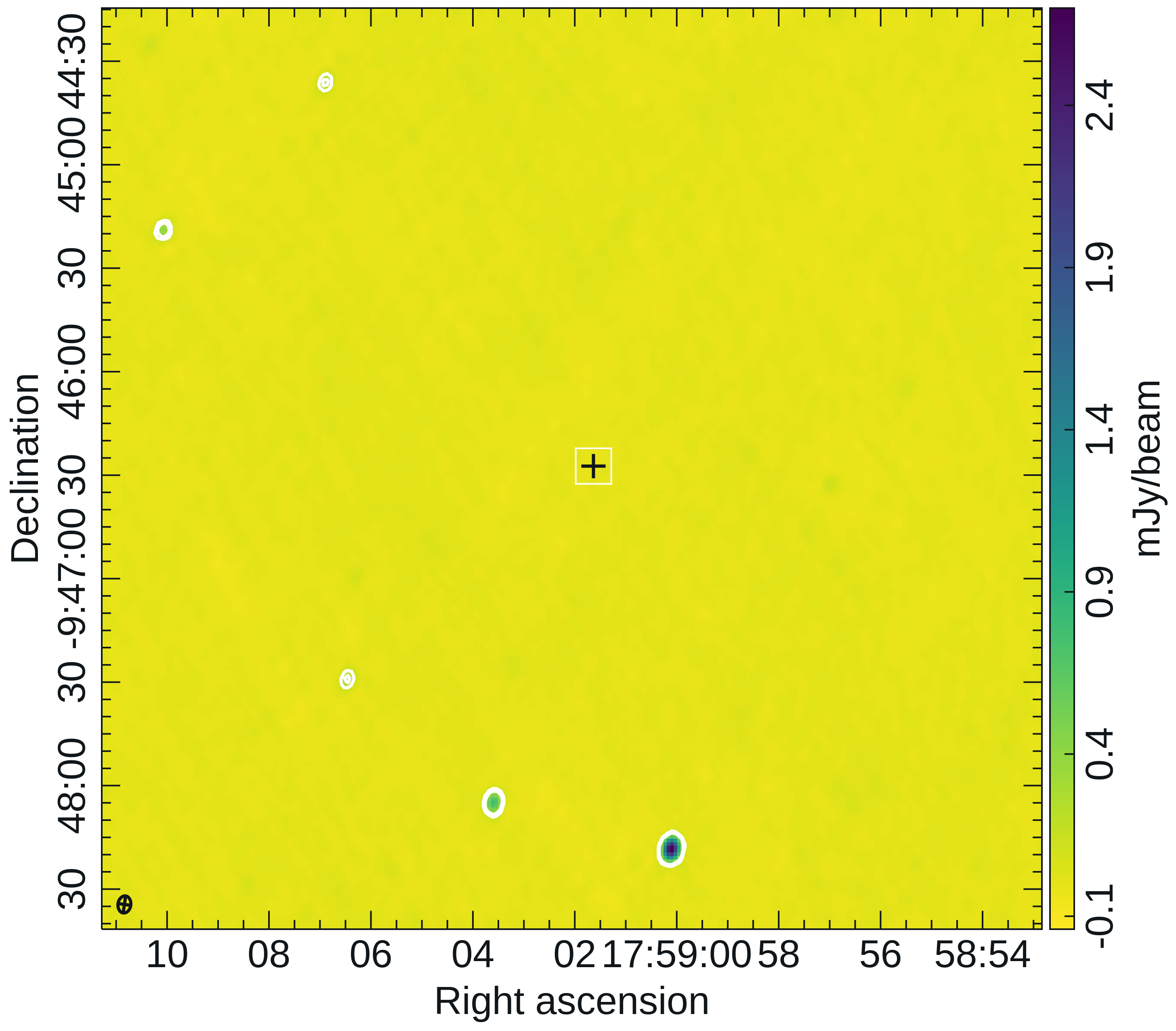}{VLA band L (1-2 GHz): $\nu$ Oph}
\end{tabular}%
}
\vspace{-0.25\baselineskip}
\caption{Stokes I maps of each source, combining all data for a given instrument and band. The colour bar spans from $-3\sigma_{I}$ to the maximum peak flux density of the field sources. The white contours mark the $-3, 3, 5\,\sigma_{I}$ levels. The clean beam is shown in the lower left corner of each image.}
\label{fig:maps}
\end{figure*}

Fig. \ref{fig:maps} shows the Stokes I clean maps, obtained by integrating over all the epochs for each target respectively. In all observations we found other bright ($\gtrsim$ mJy in Stokes I) sources in the field of view, which can affect the image quality. Note that, compared to VLA, uGMRT images tend to show much more artificial structures, coming probably from sidelobe contamination. 
For instance, a radio source close to the HD 153557 system (1'5" away, not coincident with the wide companion HD 153525) has an integrated flux of $\sim 30$ mJy in Stokes I. Close to the BD 004475 system, a binary radio source is seen at only $\approx$ 23", with fluxes of $\sim 4.7$ mJy (VLA Band L, 1-2 GHz) and $13$ mJy (uGMRT Band 4, 0.55-0.75 GHz) in Stokes I. Similarly, a bright radio source is seen close to 2M0122-2439, likely the counterpart of the eROSITA source 1eRASS J012244.4-244028. Stokes V maps (not shown) are instead characterized by the much less background source contamination, i.e. often a lower noise (see Table \ref{tab:obs}).


\bsp	
\label{lastpage}
\end{document}